\documentclass[11pt,a4paper]{article}
\usepackage[T1]{fontenc}
\usepackage{lmodern}
\usepackage{amsmath,amssymb,amsfonts,mathtools}
\usepackage{bm}
\usepackage{booktabs,array,longtable}
\usepackage{graphicx}
\graphicspath{{code/figures/}}
\usepackage[margin=2.35cm]{geometry}
\usepackage[colorlinks=true,citecolor=blue!55!black,linkcolor=blue!55!black,urlcolor=blue!55!black]{hyperref}
\usepackage[table,dvipsnames]{xcolor}
\usepackage{microtype}
\usepackage{enumitem}
\numberwithin{equation}{section}
\allowdisplaybreaks[2]
\newcommand{\dd}{\mathrm d}
\newcommand{\pd}{\partial}
\newcommand{\Ric}{\operatorname{Ric}}
\newcommand{\Tr}{\operatorname{Tr}}

\newcommand{\e}{\mathrm e}
\newcommand{\cE}{\mathcal E}
\newcommand{\cB}{\mathcal B}
\newcommand{\cD}{\mathcal D}
\newcommand{\cV}{\mathcal V}

\title{Holographic Homogeneous Black Branes from Anisotropic Scaling}

\author{
    Markus A. G. Amano\\
    \small Department of Innovative Electrical \& Electronic Engineering\\
    \small National Institute of Technology, Oyama College\\
    \small Oyama, Tochigi 323-0806, Japan\\
    \small \href{mailto:markus@oyama-ct.ac.jp}{markus@oyama-ct.ac.jp}
}

\date{November 16, 2026}

\begin{document}

\maketitle

\begin{abstract}
We construct homogeneous anisotropic black branes in Einstein gravity and Einstein--Maxwell--dilaton theory in arbitrary spacetime dimensions. By imposing compatibility between radial scaling and the spatial Lie algebra, we reduce the field equations to radial equations and algebraic constraints on the scaling exponents, curvature, and matter fields. We construct neutral, electric, magnetic, and dyonic solutions, including explicit five-dimensional Bianchi geometries and higher-dimensional families supported by Ricci nilsolitons.
We investigate the thermodynamics of these solutions, obtaining their Euclidean action, mass, and entropy scaling. We also study the dynamics of a massive probe scalar, deriving an asymptotic mass bound and an exact purely damped quasinormal spectrum for a particular class of scaling geometries. Numerical calculations further explore the mass, dynamical exponent, and momentum dependence of scalar relaxation in Nil, Solv, and Bianchi type VI geometries. Our results establish a framework for studying thermodynamic and dynamical properties of homogeneous anisotropic scaling geometries that are generally not asymptotically locally anti-de Sitter.
\end{abstract}


\section{Introduction}
\label{sec:intro}

Holography provides a useful way to study strongly coupled systems in which perturbative methods are not available \cite{Anderson:2004yi}.
In its simplest form, the gravitational description is given by an asymptotically anti-de Sitter (AdS) spacetime with a planar boundary.
However, spatial anisotropy and nontrivial scaling behavior arise in a variety of physical systems, including strongly correlated matter.
Examples include electronic nematic phases, where rotational symmetry is spontaneously broken and the low-energy excitations exhibit direction-dependent behavior \cite{Oganesyan:2001sob,Metlitski:2010pd}, and quantum critical metals near spin-density-wave ordering transitions \cite{Metlitski:2010vm}.
Lifshitz critical points can exhibit anisotropic scaling and sensitivity to spatial anisotropy \cite{Pleimling:2001zz,Diehl:2003zz}, while semi-Dirac spectra with linear and quadratic dispersion along different directions have been predicted in oxide nanostructures and modeled theoretically \cite{Pardo:2009zz,Banerjee:2009zz}.
In a separate setting, non-Hermitian models can exhibit anisotropic, system-size-dependent scaling of localization lengths \cite{Ou:2025asl}.
Spatially modulated phases also appear in strongly correlated electron systems, including electronic stripe and charge-density-wave phases \cite{Sun:2008fl,Fradkin:2015svg,Delacretaz:2017zxd}, as well as in chiral magnets supporting helical and spiral magnetic configurations \cite{Hongo:2020xaw,Amari:2023bmx}.
Related spatially modulated structures arise in chiral soliton lattices under strong magnetic fields \cite{Brauner:2016pko}.
Although their microscopic mechanisms differ, these examples motivate the study of gravitational backgrounds with homogeneous but anisotropic spatial geometries and nontrivial scaling exponents.
In particular, Einstein--Maxwell--dilaton (EMD) gravity provides a tractable framework for investigating the thermodynamic and dynamical properties of strongly coupled systems exhibiting such behavior.

Lifshitz and hyperscaling-violating backgrounds are familiar examples of gravitational geometries with nonrelativistic scaling \cite{Kachru:2008yh,Dong:2012se,Tarrio:2011de,Baggioli:2024vza}; anisotropic spacetimes were also introduced in Ho\v{r}ava gravity \cite{Horava:2008ih,Horava:2009uw} and as anisotropic conformal infinity \cite{Horava:2010vho,Oh:2021cwm}. The holographic dictionary for Lifshitz-type backgrounds is developed in \cite{Ross:2011gu,Chemissany:2012du,Papadimitriou:2014lia,Cong:2024pvs}; see \cite{Taylor:2015glc} for a review of Lifshitz holography.
More general anisotropic spacetimes can be obtained if the spatial sections are homogeneous but not necessarily isotropic.
In five bulk dimensions, the three-dimensional spatial geometries can be classified in terms of Bianchi groups, closely related to the Thurston geometries \cite{Gegenberg:2002xj,Popovych:2003xb}. Homogeneous spacetimes carrying Lifshitz symmetry are classified by algebraic methods in \cite{Figueroa-OFarrill:2022kcd}.
These provide a convenient setting in which one can preserve homogeneity without requiring ordinary spatial translation invariance \cite{Iizuka:2012iv,Iizuka:2012wt}.
Such geometries have also been used to describe helical lattices and spatially modulated states holographically \cite{Donos:2012js,Donos:2013woa,Donos:2014oha,Donos:2013eha,Donos:2014yya}.

Homogeneous black holes with Nil and Solv horizons were constructed in Einstein gravity by Cadeau and Woolgar \cite{Cadeau:2000tj}.
These were subsequently generalized to scaling and hyperscaling-violating solutions \cite{Hassaine:2015ifa}.
Higher-dimensional examples based on Einstein solvmanifolds and Ricci nilsolitons were considered in \cite{Hervik:2003vx,Hervik:2007zz}; more recent work constructs black holes with nilpotent horizons in higher dimensions \cite{Figueroa:2024utw}.
In a previous paper \cite{Amano:2026ayj}, we constructed a family of five-dimensional vacuum black branes with a negative cosmological constant and Bianchi type VI spatial geometry, characterized by a parameter $h$.
We also obtained a Ricci-flat branch when the cosmological constant vanishes, which exhibits hyperscaling violation.
The Bianchi VI$_h$ solutions include the known Solv case as a special point and show that the Einstein equations can admit nontrivial homogeneous anisotropic scaling even in the absence of matter fields.

A dilaton and electromagnetic fields enlarge the set of allowed scaling geometries. An electric field can change the dynamical exponent, while a magnetic field distinguishes spatial planes and must satisfy the Maxwell equation and Bianchi identity on the chosen homogeneous space. Matter-supported black holes with Thurston horizon geometries, including dyonic solutions, have been studied in \cite{Bravo-Gaete:2017nkp,Arias:2017yqj,Arias:2018mqn}. Magnetic backgrounds also provide a setting for anisotropic holographic transport, and other anisotropic holographic backgrounds are constructed from axion or dilaton deformations and from top-down setups \cite{Donos:2016zpf,Giataganas:2017koz,Penin:2017lqt,Gursoy:2020kjd}; anisotropic critical points from holography are studied in \cite{Giataganas:2025ing}. The distinction between a field that supports the asymptotic geometry and an independently variable charge is essential here: the explicit EMD families considered below belong to the former class.

We formulate the neutral and EMD equations in arbitrary spacetime dimension using a diagonal metric in a left-invariant spatial coframe. Let $r$ denote the radial coordinate and $\alpha_i$ the scaling exponent of the $i$th spatial direction; the corresponding spatial metric coefficient is proportional to $r^{2\alpha_i}$. The spatial Lie algebra restricts the radial exponents through scaling compatibility and the mixed Einstein equations. The remaining equations determine the intrinsic Ricci tensor, blackening function, and matter amplitudes. For the neutral solutions, we distinguish a branch with radially constant intrinsic spatial curvature and a negative cosmological constant from one with radially varying intrinsic curvature and a vanishing cosmological constant, and explain the relation to algebraic Ricci solitons. The matter equations are organized into electric, magnetic, and dyonic sectors, retaining the geometric restrictions on homogeneous magnetic fields. The examples include five-dimensional Bianchi geometries, a dyonic Nil solution with a single Abelian gauge field, flat higher-dimensional dyonic solutions, and a construction that adjoins a magnetic two-plane to a Ricci nilsoliton.

Let $r_h>0$ be the horizon radius and $f_0>0$ the asymptotic value of the blackening function. For a positive exponent $\Delta$ equal to the dynamical exponent plus the sum of the spatial scaling exponents, we consider solutions with blackening function
\begin{equation}
    \label{eq:intro_blackening}
    f(r)=f_0\left[1-\left(\frac{r_h}{r}\right)^\Delta\right].
\end{equation}
The neutral planar AdS black brane with $n$ spatial directions and AdS radius $L$ has $f_0=L^{-2}$ and $\Delta=n+1$ in the metric convention used here \cite{Birmingham:1998nr}. Thus, \eqref{eq:intro_blackening} generalizes its blackening function by allowing a different exponent and normalization, without requiring the geometry to be asymptotically AdS.
For these solutions, we calculate the temperature, entropy, and Euclidean action with the supporting sources held fixed. Including the electric Legendre term gives a fixed-flux free energy relative to the corresponding horizonless scaling background. The resulting energy agrees with the fixed-source Hamiltonian variation. In this family, the entropy scales as a power of temperature, with exponent given by the sum of the spatial scaling exponents divided by the dynamical exponent. The heat capacity is positive when both the sum of the spatial exponents and the dynamical exponent are positive. This calculation determines relative thermodynamic quantities within the specified family; a variational problem with arbitrary boundary sources requires a more general renormalization prescription \cite{Papadimitriou:2010as}.

We then examine a massive probe scalar to investigate how fluctuations propagate and relax on these anisotropic branes. The momentum dependence of the scalar quasinormal frequencies provides dispersion relations and a first view of the dynamical response relevant to transport. It does not, by itself, determine hydrodynamic modes or transport coefficients of the dual theory. The scalar also provides a tractable probe of stability against disturbances in this sector.

In the spatially homogeneous sector, the asymptotic equation gives a mass bound analogous to the Breitenlohner--Freedman bound, and the quasinormal frequencies scale linearly with temperature at fixed background parameters. When the dynamical exponent equals the sum of the spatial scaling exponents and the blackening function has no logarithmic term, the radial equation becomes hypergeometric. For masses satisfying the asymptotic bound, it yields an exact, purely damped spectrum under source-free boundary conditions. We use this result as a benchmark for a pseudospectral calculation of the electric Nil, Solv, and Bianchi VI$_h$ families. The numerical study follows the dependence on probe mass and dynamical exponent and examines finite-momentum sectors of Nil and Solv. Resolution comparisons and a separate shooting calculation assess the accuracy and limitations of the verified modes (Appendix~\ref{app:numerics}). The exact spectrum is consistent with stability in the homogeneous scalar sector under the stated boundary conditions, but the numerical selection keeps only damped modes and cannot rule out growing modes. Neither calculation establishes stability of the coupled metric, gauge-field, and dilaton system.

The diagonal metric ansatz with these radial scaling coefficients restricts the class of backgrounds considered here. The solutions are generally not asymptotically locally AdS, although they may serve as scaling regions in more general holographic geometries. Their explicit form allows us to separate the restrictions imposed by spatial geometry from those imposed by matter and to study how both enter the thermodynamics and scalar response.

The organization of the paper is as follows.
In Sec.~\ref{sec:neutral} we derive the neutral Einstein equations and examine the two principal anisotropic branches, including their relation to earlier homogeneous black holes.
In Sec.~\ref{sec:matter} we introduce a dilaton and an Abelian gauge field, and discuss the electric, magnetic, and dyonic cases.
Section~\ref{sec:examples} contains representative five-dimensional and higher-dimensional examples.
In Sec.~\ref{sec:thermo} we derive the horizon thermodynamics and evaluate the fixed-flux Euclidean on-shell action.
Section~\ref{sec:probe_scalar} studies massive probe scalar perturbations, including their asymptotic falloffs, quasinormal boundary conditions, an analytic spectrum when the dynamical exponent equals the sum of the spatial scaling exponents, and a pseudospectral computation of the spectra with its validation (Appendix~\ref{app:numerics}).
We conclude in Sec.~\ref{sec:discussion} with a summary and possible extensions.
Appendix~\ref{app:radial} summarizes the radial integration, including the case where a source term has the same radial power as a homogeneous mode, Appendix~\ref{app:mass} gives the fixed-source Hamiltonian variation underlying the relative mass, and Appendix~\ref{app:numerics} documents the convergence and reliability checks of the numerical spectra.

\section{Neutral homogeneous anisotropic black branes}
\label{sec:neutral}

\subsection{Einstein gravity and homogeneous geometry}

We first consider Einstein gravity with a cosmological constant in $D:=n+2$ dimensions. We work in units $8\pi G_N=1$. The action is
\begin{equation}
    \label{eq:neutral_action}
    I:=\frac12\int \dd^{n+2}x\,\sqrt{-g}\,(R-2\Lambda)+I_{\mathrm{bdy}},
\end{equation}
where $I_{\mathrm{bdy}}$ contains the boundary terms required for a suitable variational principle.
The equations of motion are
\begin{equation}
    \label{eq:neutral_ee}
    R_{\mu\nu}-\frac12 Rg_{\mu\nu}+\Lambda g_{\mu\nu}=0,
    \qquad R_{\mu\nu}=\frac{2\Lambda}{n}g_{\mu\nu}.
\end{equation}

Let $G$ be an $n$-dimensional Lie group with Lie algebra $\mathfrak g$\footnote{
    More general homogeneous spatial manifolds can also be considered; here we restrict attention to Lie groups.
} with a left-invariant coframe $\theta^i$ satisfying
\begin{equation}
    \label{eq:maurer_cartan}
    \dd\theta^k=-\frac12 c_{ij}{}^k\,\theta^i\wedge\theta^j.
\end{equation}
The structure constants $c_{ij}{}^k$ are constant in this invariant coframe. The metric
\begin{equation}
    \label{eq:general_metric}
    \dd s^2=-G_t(r)\dd t^2+G_r(r)\dd r^2+
    \sum_{i=1}^n G_i(r)(\theta^i)^2
\end{equation}
is homogeneous on the spatial slices of constant $r$, but not necessarily isotropic. Unless stated otherwise, frame indices $t,r,i,j,k,\ldots$ denote components in the orthonormal frame of the full metric. Thus, the field equations reduce to ordinary differential equations in the radial coordinate $r$. We assume that the intrinsic Ricci tensor of the reference spatial metric can be made diagonal in the chosen coframe.

Following the constructions in \cite{Hassaine:2015ifa, Cadeau:2000tj, Amano:2026ayj}, we choose the following diagonal metric ansatz
\begin{equation}
    \label{eq:metric_ansatz}
    \dd s^2=-r^{2z}f(r)\dd t^2+\frac{\dd r^2}{r^2f(r)}+
    \sum_{i=1}^n b_i^2r^{2\alpha_i}(\theta^i)^2.
\end{equation}
The parameters $z$ and $\alpha_i$ control the radial scaling of time and the spatial directions, respectively, while the constants $b_i>0$ determine the spatial curvature scales. The function $f(r)$ is the blackening factor. Let $X_i$ denote the spatial vector fields dual to the invariant coframe, $\theta^i(X_j)=\delta^i{}_j$. In the exterior region $f>0$, the orthonormal frame vectors of the full metric are
\begin{equation}
    \label{eq:full_orthonormal_frame}
    e_t:=\frac{1}{r^z\sqrt{f}}\,\partial_t,\qquad
    e_r:=r\sqrt{f}\,\partial_r,\qquad
    e_i:=\frac{1}{b_i r^{\alpha_i}}X_i.
\end{equation}
Thus $g(e_t,e_t)=-1$ and $g(e_r,e_r)=g(e_i,e_i)=1$ (no sum). For notational simplicity we define
\begin{equation}
    \label{eq:SQDelta}
    S:=\sum_{i=1}^n\alpha_i,\qquad
    Q:=\sum_{i=1}^n\alpha_i^2,\qquad
    \Delta:=z+S,\qquad
    \cD:=r\frac{\dd}{\dd r}.
\end{equation}
In most of this paper we take $r>r_h$, $z>0$, $S>0$, and $f>0$ outside the horizon where $r_h$ is the horizon radius. In addition to constraints from the Einstein equations, we will also impose additional constraints on $\alpha_i$.

\subsection{Scaling compatibility and mixed Einstein equations}

The metric ansatz \eqref{eq:metric_ansatz} assigns a radial scaling exponent to each spatial direction. This is straightforward for a flat horizon, but a non-Abelian homogeneous space has an additional feature: two spatial directions can be mixed by the geometry itself. To obtain solutions of the form \eqref{eq:metric_ansatz}, we impose
\begin{equation}
    \label{eq:kappa}
    c_{ij}{}^k\neq0\quad\Longrightarrow\quad
    \alpha_i+\alpha_j-\alpha_k=\kappa.
\end{equation}
As a consequence the connection associated with the spatial geometry then scales as $r^{-\kappa}$, and its intrinsic Ricci tensor carries a common factor $r^{-2\kappa}$. In particular, $\kappa=0$ means that the intrinsic curvature contributes at the same radial order as the cosmological constant. This is the branch that admits the asymptotically constant blackening factors studied below. We write the reference spatial metric and its Ricci eigenvalues as
\begin{equation}
    \label{eq:ricci_ref}
    h_b:=\sum_i b_i^2(\theta^i)^2,\qquad
    \Ric(h_b)_{\hat i\hat j}=:R_{\hat i\hat j}.
\end{equation}
Here hatted indices refer to the orthonormal frame $\hat e_i:=b_i^{-1}X_i$ of $h_b$.

For $\kappa=0$, \eqref{eq:kappa} has a simple scaling interpretation. A radial dilatation $r\to\ell r$ can be accompanied by a transformation of the invariant one-forms $\theta^i\to\ell^{-\alpha_i}\theta^i$, so that the zero-temperature metric retains its form. The condition on the structure constants ensures that this transformation preserves the relations between the one-forms. Equivalently, the map 
\begin{equation}
    \label{eq:A_operator}
    A(X_i):=\alpha_i X_i
\end{equation}
on the spatial Lie algebra satisfies
\begin{equation}
    A([X_i,X_j])=[AX_i,X_j]+[X_i,AX_j].
\end{equation}
This is called a derivation; it need not be an \emph{inner} derivation. An associated dilatation generator $X_A$ can be adjoined to $\mathfrak g$, with its action defined by the bracket $[X_A,X_i]:=A(X_i)=\alpha_i X_i$. For $\kappa\neq0$, the corresponding extension is instead generated by the derivation $A-\kappa\mathbf1$. Thus the anisotropic radial scaling must preserve the homogeneous spatial geometry. Such structures also appear in earlier homogeneous Einstein constructions~\cite{Hervik:2003vx,Hervik:2007zz}.

Nevertheless, the parametrization of the solutions carries a residual redundancy. Performing the radial diffeomorphism $r\to r^\gamma$, together with $t\to t/\gamma$ so that the ansatz keeps its canonical form, rescales the parameters as
\begin{equation}
    \label{eq:radial_rescale}
    (z,\alpha_i,S,\Delta,\sigma)\to\frac{1}{\gamma}\,(z,\alpha_i,S,\Delta,\sigma),
    \qquad
    f_0\to\gamma^2 f_0,\qquad
    r_h\to r_h^\gamma,
\end{equation}
while the constants $b_i$, $\phi_0$, $\cE$, $\cB$, and $\Lambda$ are unchanged. The overall scale of the exponent vector is therefore a coordinate artifact. If $\kappa\neq0$, this freedom can be used to set $\kappa=1$. If $\kappa=0$, the value of $\kappa$ cannot be changed, and the freedom is instead used to fix the overall scale of the exponents, which we denote by $a$, to unity. The cases $\kappa=0$ and $\kappa\neq0$ therefore define distinct scaling branches.

The mixed radial--spatial Einstein equations impose a further compatibility condition. The diagonal metric and radial matter fields carry no momentum flux between the radial direction and a spatial direction. The geometry must therefore produce no such off-diagonal curvature either. For the ansatz~\eqref{eq:metric_ansatz}, this gives
\begin{equation}
    \label{eq:ri_constraint}
    R_{ri}=\frac{\sqrt{f}}{b_i r^{\alpha_i}}
    \sum_j c_{ij}{}^j(\alpha_j-\alpha_i)=0
    \quad\Longrightarrow\quad
    \sum_j c_{ij}{}^j(\alpha_j-\alpha_i)=0.
\end{equation}
This condition is automatic for an Abelian group and for isotropic spatial scaling, but can be restrictive for a non-Abelian geometry. For example, in Bianchi VI$_h$ it gives $\alpha_1+h\alpha_2=(1+h)\alpha_3$, or $\alpha_1=-h\alpha_2$ for $\alpha_3=0$~\cite{Amano:2026ayj}.

\subsection{Trace-reversed Einstein equations}

Using an orthonormal frame, the nontrivial diagonal components of the Ricci tensor are
\begin{align}
    R_{tt}={}&\frac12r^2f''+\frac12(1+3z+S)rf'+z(z+S)f,\label{eq:Rtt}\\
    R_{rr}={}&-\frac12r^2f''-\frac12(1+3z+S)rf'-(z^2+Q)f,\label{eq:Rrr}\\
    R_{ii}={}&R_{\hat i \hat i} r^{-2\kappa}-\alpha_i\left[rf'+(z+S)f\right].\label{eq:Rii}
\end{align}
In our orthonormal signature the trace-reversed vacuum equations are 
\begin{equation}
    R_{tt}=-2\Lambda/n,\quad R_{rr}=2\Lambda/n,\quad R_{ii}=2\Lambda/n.
\end{equation}
Adding the first two equations immediately yields
\begin{equation}
    \label{eq:vacuum_z}
    zS=Q.
\end{equation}
This equation is useful because it fixes the time scaling exponent before solving for $f$. For nonzero spatial scaling, $S\neq0$ and $z=Q/S$.

Substituting \eqref{eq:vacuum_z} into the temporal equation gives
\begin{equation}
    \label{eq:vacuum_f_op}
    (\cD+2z)(\cD+\Delta)f=-\frac{4\Lambda}{n}.
\end{equation}
For $Q\neq S^2$, the solution is
\begin{equation}
    \label{eq:vacuum_f_generic}
    f(r)=-\frac{2\Lambda}{nz\Delta}
    +\frac{C_1}{r^{2z}}+\frac{C_2}{r^\Delta}.
\end{equation}
The homogeneous radial modes have powers $2z$ and $\Delta$, while the cosmological constant contributes a constant particular solution. The spatial equations further constrain the integration constants.

The spatial equations can be written as
\begin{equation}
\label{eq:vacuum_spatial_exact}
\begin{aligned}
        \alpha_i(\cD+\Delta)f &=R_{\hat i \hat i} r^{-2\kappa}-\frac{2\Lambda}{n} \\
        \implies R_{\hat i \hat i} r^{-2\kappa}-\alpha_i(S-z)C_1r^{-2z}
        &=\frac{2\Lambda}{n}\left(1-\frac{\alpha_i}{z}\right).\\
\end{aligned}
\end{equation}
Note that $(\cD+\Delta)\,r^{-\Delta}=0$, so only the $C_1$ mode
$\propto r^{-2z}$ contributes $(\cD+\Delta)(r^{-2z})=(S-z)r^{-2z}$
and thus appears in \eqref{eq:vacuum_spatial_exact}.

For $S\neq0$, the case $Q=S^2$ corresponds to a repeated radial root. Here $z=S$ and
\begin{align}
    f(r)&=-\frac{\Lambda}{nS^2}+r^{-2S}\bigl(C_1+C_2\log r\bigr),\label{eq:vacuum_repeated}\\
    \implies R_{\hat i \hat i} r^{-2\kappa}-\alpha_i C_2r^{-2S}
        &=\frac{2\Lambda}{n}\left(1-\frac{\alpha_i}{S}\right).\label{eq:vacuum_repeated_spatial}
\end{align}
The logarithmic term cannot be obtained by substituting $z=S$ into the distinct-root solution. As shown below, the spatial equations eliminate it when $\kappa=0$ and $S\neq0$, but it may appear in special $\Lambda=0$ geometries.

\subsection{Neutral solution branches}
\label{sec:neutral_solutions}

\paragraph{Anisotropic solutions for $\kappa=0$ and $\Lambda<0$}

We first take $\kappa=0$ and $\Lambda<0$. For a genuinely anisotropic solution with $S\neq0$, the spatial equations fix $C_1=0$ in the case where the radial equation has distinct roots, $Q\neq S^2$, and require
\begin{equation}
    \label{eq:neutral_ricci_constraint}
    R_{\hat i \hat i}=\frac{2\Lambda}{n}\left(1-\frac{\alpha_i}{z}\right).
\end{equation}
For the repeated-root case $Q=S^2$ with $S\neq0$, \eqref{eq:vacuum_z} gives $z=S$ and $\Delta=2S$. Setting $\kappa=0$ in \eqref{eq:vacuum_repeated_spatial} shows that $\alpha_i C_2=0$ for every $i$, since the remaining terms are independent of $r$. At least one $\alpha_i$ is nonzero, so $C_2=0$ and the logarithmic mode is excluded. The remaining spatial condition is again \eqref{eq:neutral_ricci_constraint}, now with $z=S$, and \eqref{eq:vacuum_repeated} reduces to
\begin{equation}
    \label{eq:neutral_repeated_f}
    f(r)=-\frac{\Lambda}{nS^2}+\frac{C_1}{r^{2S}}.
\end{equation}
Here $C_1$ multiplies the surviving repeated-root mode; it is not the coefficient set to zero in the distinct-root case. The constant part of $f$ is $f_0=-\Lambda/(nS^2)>0$. For $S>0$, choosing $C_1<0$ gives a horizon at $r_h=(-C_1/f_0)^{1/(2S)}$.

Thus, in either the distinct-root or repeated-root case, the blackening function of the horizon solution can be written as
\begin{equation}
    \label{eq:neutral_f}
    f(r)=f_0\left[1-\left(\frac{r_h}{r}\right)^\Delta\right], \qquad
    f_0=-\frac{2\Lambda}{nz\Delta}.
\end{equation}
For $z>0$, $S>0$, the normalization $f_0$ is positive and $f$ has a simple zero at $r=r_h$. In particular, the repeated-root case has $\Delta=2S$ and $f_0=-\Lambda/(nS^2)$. The $h=0$ member of the Bianchi VI$_h$ family in Sec.~\ref{sec:examples_neutral} realizes this case.

The spatial equations have a useful interpretation beyond the algebraic restriction on $R_{\hat i \hat i}$. In vacuum there is no anisotropic matter stress to distinguish the spatial directions. Nevertheless, the intrinsic curvature of the homogeneous slices is direction dependent and can balance their different radial expansion rates. \eqref{eq:neutral_ricci_constraint} specifies precisely how much curvature each direction must carry. In particular, when $\Lambda<0$, a direction with $\alpha_i>z$ requires a positive intrinsic Ricci eigenvalue, while directions with $\alpha_i<z$ require a negative one. The Nil example in section \ref{sec:examples_neutral} shows this clearly.

In geometric terms, using the scaling map \eqref{eq:A_operator}, the same condition reads
\begin{equation}
    \label{eq:ricci_soliton}
    \Ric(h_b)=\frac{2\Lambda}{n}\left(\mathbf1-\frac{1}{z}A\right).
\end{equation}
This is a form of algebraic Ricci soliton condition, as in \cite{Hervik:2007zz,Lauret:2011jo}. For the present calculation, it provides a convenient way to identify homogeneous spatial geometries whose directional curvature can support an anisotropic black brane without additional matter fields.

\paragraph{Isotropic Einstein horizons}

The isotropic Einstein horizon case is a separate specialization of the metric ansatz \eqref{eq:metric_ansatz}. Take $n>1$ and equal nonzero spatial exponents $\alpha_i=a>0$. The neutral constraint \eqref{eq:vacuum_z} then gives $z=a$. For a non-Abelian spatial coframe, the scaling condition \eqref{eq:kappa} requires $\kappa=a$, so a curved isotropic horizon does not belong to the $\kappa=0$ branch above. For an Abelian coframe the condition is vacuous, and the spatial metric is flat.

Let the reference spatial metric be Einstein, with $\Ric(h_b)=(n-1)k\,h_b$, where $k$ is its normalized Ricci curvature. For the Lie-group ansatz, the spatial equations give $k=a^2C_1$; in the Abelian case both vanish. Writing the remaining integration constant as $\mu:=-a^2C_2$, and defining $R:=r^a$ and $\tilde t:=t/a$, the metric takes the familiar form
\begin{equation}
    \label{eq:einstein_horizon}
    \dd s^2=-F(R)\dd{\tilde t}^2+\frac{\dd R^2}{F(R)}+R^2h_b,\quad
    F(R)=k-\frac{\mu}{R^{n-1}}-\frac{2\Lambda R^2}{n(n+1)}.
\end{equation}
For $\Lambda<0$ and suitable $\mu$, this includes topological AdS black holes with flat, spherical or hyperbolic horizons \cite{Banados:1992wn,Mann:1997jb,Vanzo:1997gw,Birmingham:1998nr}; the stability of these black holes is analyzed in \cite{Birmingham:2007yv}. The expression also applies to Einstein horizons beyond the Lie-group coframes assumed in the rest of this paper.

\paragraph{Vanishing cosmological constant and $\kappa\neq0$}

Another possibility is obtained by setting $\Lambda=0$. If $Q\neq S^2$, the temporal equation becomes
\begin{equation}
    f(r)=C_1r^{-2z}+C_2r^{-\Delta}.
\end{equation}
For nonzero intrinsic spatial curvature, the spatial Einstein equations require
\begin{equation}
    \label{eq:neutral_kappane0}
    \kappa=z,\qquad R_{\hat i \hat i}=\alpha_i(S-z)C_1.
\end{equation}
This gives an anisotropic Ricci-flat spacetime even though the individual spatial slices are curved. The spatial curvature is balanced by the radial warp factors.
These $\Lambda=0$ solutions are hyperscaling-violating geometries: the spatial warp factors carry unequal exponents $\alpha_i$, so the effective number of spatial dimensions differs from $n$ even though the full spacetime is Ricci flat. Vacuum black holes of this type were constructed for the Nil geometry in \cite{Hassaine:2015ifa} and generalized to the Bianchi VI$_h$ family in \cite{Amano:2026ayj}; hyperscaling-violating geometries were introduced holographically in \cite{Charmousis:2010zz,Dong:2012se}.
In the case of a repeated radial root, $Q=S^2$, the corresponding nonflat branch is
\begin{equation}
    \label{eq:neutral_log_branch}
    \kappa=S,\quad R_{\hat i \hat i}=\alpha_i C_2,\quad
    f(r)=r^{-2S}(C_1+C_2\log r).
\end{equation}

\paragraph{Zero spatial scaling and other special cases}

When $S=0$, \eqref{eq:vacuum_z} requires $Q=0$ for real neutral exponents. Hence $\alpha_i=0$ in every spatial direction. The spatial metric must be Einstein, with $R_{\hat i \hat i}=2\Lambda/n$. Using $\rho:=\log r$, the metric can be written as a two-dimensional constant curvature factor times $h_b$,
\begin{equation}
    \label{eq:product_neutral}
    \dd s^2=-U(\rho)\dd t^2+\frac{\dd\rho^2}{U(\rho)}+h_b,
    \qquad U(\rho)=-\frac{2\Lambda}{n}\rho^2+c_1\rho+c_2.
\end{equation}
For $\Lambda<0$, this includes AdS$_2$ product geometries. Such branches are related to generalized (anti-)Nariai constructions \cite{Nariai:1950na,Nariai:1999iok,Ishihara:1983au,Cardoso:2004uz}. Appendix~\ref{app:radial} gives the radial integration rule for the cases of distinct and repeated roots, including the case where a source term has the same radial power as a homogeneous mode. The cases where $S$ or $z$ vanishes, or where $Q=S^2$, must be treated separately, because the general formulas divide by $S$, $z$, or $Q-S^2$ and can otherwise remove valid solutions.

\section{Einstein--Maxwell--dilaton black branes}
\label{sec:matter}

In the preceding section we saw that homogeneous spatial curvature can support anisotropic scaling even in vacuum. We now extend the construction by adding a $U(1)$ gauge field and a dilaton. The electromagnetic field changes the spatial Einstein equations and, consequently, the conditions on the scaling exponents. Our main interest is to understand which homogeneous geometries admit electric and magnetic fields without losing the cohomogeneity-one form of the metric.

\subsection{Action and equations of motion}

We take the Einstein--Maxwell--dilaton (EMD) action
\begin{equation}
    \label{eq:emd_action}
    I:=\frac12\int \dd^{n+2}x\sqrt{-g}\left[R-2\Lambda-(\partial\phi)^2-\frac12\e^{\lambda\phi}F_{\mu\nu}F^{\mu\nu}\right]+I_{\mathrm{bdy}},\qquad F:=\dd A.
\end{equation}
Here $\lambda$ is a constant coupling and $\phi$ is a real scalar. We use the convention where $8\pi G_N=1$. The field equations, in trace-reversed form, are
\begin{align}
    R_{\mu\nu}&=\frac{2\Lambda}{n}g_{\mu\nu}+\pd_\mu\phi\pd_\nu\phi+\e^{\lambda\phi}\left(F_{\mu\rho}F_{\nu}{}^\rho-\frac{1}{2n}g_{\mu\nu}F^2\right),\label{eq:emd_einstein}\\
    \nabla_\mu(\e^{\lambda\phi}F^{\mu\nu})&=0,\qquad \dd F=0,\label{eq:emd_maxwell}\\
    \Box\phi&=\frac\lambda4\e^{\lambda\phi}F^2.\label{eq:emd_scalar}
\end{align}
We continue to use the metric~\eqref{eq:metric_ansatz}, and set
\begin{equation}
    \label{eq:scalar_ansatz}
    \phi(r)=\phi_0+\sigma\log r.
\end{equation}

\subsection{Purely electric solutions}
\label{sec:electric}

We first take a homogeneous electric field,
\begin{equation}
    A=A_t(r)\dd t,\qquad F=E(r)\dd r\wedge\dd t.
\end{equation}
The Maxwell equation is readily integrated:
\begin{equation}
    \label{eq:elect_maxwell_first}
    \frac{\dd}{\dd r}\left(r^{S+1-z+\lambda\sigma}E\right)=0,
    \qquad E(r)=q\e^{-\lambda\phi_0}r^{z-S-1-\lambda\sigma}.
\end{equation}
Defining $\cE:=q^2\e^{-\lambda\phi_0}$ and $p:=2S+\lambda\sigma$, the electric energy in a local orthonormal frame is proportional to $\cE r^{-p}$. Adding the temporal and radial Einstein equations gives
\begin{equation}
    \label{eq:emd_zS}
    zS=Q+\sigma^2.
\end{equation}
Therefore the dilaton increases the effective time exponent relative to its neutral value. For $S\neq0$ we have $z=(Q+\sigma^2)/S$. The independent equations can be written as
\begin{align}
    (\cD+2z)(\cD+\Delta)f
 &= -\frac{4\Lambda}{n}+\frac{2(n-1)}{n}\cE r^{-p},\label{eq:elect_tt}\\
 \alpha_i(\cD+\Delta)f
 &=R_{\hat i \hat i} r^{-2\kappa}-\frac{2\Lambda}{n}-\frac{\cE}{n}r^{-p},\label{eq:elect_ii}\\
 \sigma(\cD+\Delta)f
 &=-\frac{\lambda\cE}{2}r^{-p}.\label{eq:elect_phi}
\end{align}
As in the neutral case, the temporal equation determines the possible radial powers, whereas the spatial and scalar equations constrain their coefficients.

\paragraph{Nonzero cosmological constant.}
For a running dilaton, $\sigma\neq0$, and $\Lambda\neq0$, the scalar equation and the constant term of the temporal equation require a constant electric source. Hence,
\begin{equation}
    \label{eq:elect_const_source}
    p=0\Longleftrightarrow \lambda\sigma=-2S.
\end{equation}
For a curved homogeneous spatial metric, the common scaling ansatz further gives $\kappa=0$. The blackening factor then takes the form
\begin{equation}
    \label{eq:elect_blackening}
    f(r)=f_0\left[1-\left(\frac{r_h}{r}\right)^\Delta\right],\qquad
    f_0=\frac{-2\Lambda+(n-1)\cE}{nz\Delta}.
\end{equation}
The scalar and spatial equations imply
\begin{align}
    z\Delta f_0&=\frac{-2\Lambda+(n-1)\cE}{n},\\
    \sigma\Delta f_0&=-\frac{\lambda\cE}{2},\label{eq:elect_algebraic}\\
    R_{\hat i \hat i}&=\alpha_i\Delta f_0+\frac{2\Lambda+\cE}{n}.
\end{align}
Combining the first two relations with~\eqref{eq:emd_zS} and~\eqref{eq:elect_const_source} yields
\begin{equation}
    \label{eq:elect_amplitude}
    \cE=-\frac{2\Lambda\sigma^2}{nQ+\sigma^2},\qquad
    f_0=-\frac{2\Lambda S^2}{(nQ+\sigma^2)(Q+\sigma^2+S^2)}.
\end{equation}
For $\Lambda<0$, these amplitudes are positive whenever $S>0$ and $\sigma\neq0$. The electric amplitude is fixed by the scaling exponents, $\lambda$, and $\Lambda$ and supports the asymptotic geometry. It does not introduce an independent charge term in the blackening function.

\paragraph{Vanishing cosmological constant and nonconstant source powers.}
For $\Lambda=0$, an electric source need not be constant. If $p\neq2z,\Delta$, the temporal equation admits a particular solution proportional to $r^{-p}$, in addition to the homogeneous modes $r^{-2z}$ and $r^{-\Delta}$. The remaining equations constrain which of these modes survive. The cases where $p=2z$ or $p=\Delta$ give logarithmic particular solutions, and the case of a repeated radial root ($z=S$) must be integrated separately. We give a unified rule for these possibilities in Appendix~\ref{app:radial}. The spatial and scalar equations must also be satisfied.

\subsection{Homogeneous magnetic fields}
\label{sec:magnetic}

Next, we replace the electric field by a homogeneous magnetic field. A magnetic two-form can be expressed in terms of the invariant coframe as
\begin{equation}
    \label{eq:magnetic_2form}
    F_{\mathrm m}=\frac12 B_{ij}\theta^i\wedge\theta^j,\qquad B_{ij}=-B_{ji},
\end{equation}
where $B_{ij}$ are constants. Unlike a Cartesian magnetic field, the invariant two-form need not be closed for a non-Abelian spatial group. Thus the Bianchi identity and Maxwell equations impose restrictions on the allowed orientations.

A homogeneous magnetic field is not necessarily a magnetic field with constant Cartesian components. Instead, it has constant components in the invariant frame of the spatial geometry. This distinction matters because the invariant one-forms can depend on the spatial coordinates. The Bianchi identity $\dd F=0$ requires the magnetic flux to have no local sources, while the Maxwell equation requires it to be supported without an external current. Together, these requirements select the allowed magnetic planes. In five bulk dimensions, writing $M_1:=B_{23}$, $M_2:=B_{31}$, and $M_3:=B_{12}$, they take the form
\begin{align}
    \sum_{i<j}B_{ij}\dd(\theta^i\wedge\theta^j)&=0,\label{eq:magnetic_closure}\\
    \sum_i b_i^2M_i r^{2\alpha_i}\dd\theta^i&=0.\label{eq:magnetic_codiff}
\end{align}
For distinct spatial exponents, terms with different radial powers cannot cancel, so a nonzero $M_i$ requires $\dd\theta^i=0$. Equal exponents may allow cancellations. These conditions are best viewed as the magnetic analogue of the scaling-compatibility equations: they ensure that the proposed magnetic background respects the same spatial homogeneity as the metric.

For a single magnetic plane $(ij)$, let $\chi_\ell$ be one in the magnetic plane and zero in the remaining directions. Define
\begin{equation}
    \label{eq:B_def}
    \cB:=\frac{B_{ij}^2\e^{\lambda\phi_0}}{b_i^2b_j^2},\qquad
    m_B:=\lambda\sigma-2(\alpha_i+\alpha_j),\qquad
    v_\ell:=\chi_\ell-\frac1n.
\end{equation}
We assume the magnetic stress tensor is diagonal in the same frame as the spatial Ricci tensor. The field equations are
\begin{align}
    (\cD+2z)(\cD+\Delta)f
 &=-\frac{4\Lambda}{n}+\frac{2\cB}{n}r^{m_B},\label{eq:mag_tt}\\
 \alpha_\ell(\cD+\Delta)f
 &=R_{\hat\ell \hat\ell} r^{-2\kappa}-\frac{2\Lambda}{n}-v_\ell\cB r^{m_B},\label{eq:mag_ii}\\
 \sigma(\cD+\Delta)f
 &=\frac{\lambda\cB}{2}r^{m_B}.\label{eq:mag_phi}
\end{align}
The $tt+rr$ equation again gives $zS=Q+\sigma^2$.

For $\Lambda\neq0$, $\lambda\sigma\neq0$, and a curved spatial geometry, the equations require
\begin{equation}
    m_B=0,\qquad \kappa=0,\qquad \lambda\sigma=2(\alpha_i+\alpha_j).
\end{equation}
Writing $\Delta f_0$ for the common left-hand side, the remaining equations reduce to
\begin{align}\label{eq:mag_algebraic}
    f_0&=\frac{-2\Lambda+\cB}{z\Delta n},\nonumber\\
    R_{\hat\ell \hat\ell}&=\alpha_\ell \Delta f_0+\frac{2\Lambda}{n}+v_\ell\cB,\\
    \sigma &=\frac{\lambda\cB}{2\Delta f_0},\qquad zS=Q+\sigma^2.\nonumber
\end{align}
For $\Delta>0$ the blackening factor is
\begin{equation}
    \label{eq:mag_f}
    f(r)=f_0\left[1-\left(\frac{r_h}{r}\right)^\Delta\right].
\end{equation}
The important difference from the electric case is the directional magnetic stress $v_\ell\cB$. It can support anisotropic radial scaling even when the spatial group is Abelian, and it singles out the magnetic plane. For a constant dilaton and a nonzero isolated magnetic field, the scalar equation requires $\lambda=0$. The cases with $\Lambda=0$, with a magnetic field whose radial power is nonzero ($m_B\neq0$), and with a source term sharing the radial power of a homogeneous mode are summarized in Appendix~\ref{app:radial}.

\subsection{Dyonic configurations}
\label{sec:dyonic}

We now include both an electric field and a single homogeneous magnetic plane. The field strength is
\begin{equation}
    \label{eq:dyonic_ansatz}
    F=q\e^{-\lambda\phi_0}r^{z-S-1-\lambda\sigma}\dd r\wedge\dd t
    +B_{ij}\theta^i\wedge\theta^j.
\end{equation}
The magnetic constraints~\eqref{eq:magnetic_closure}--\eqref{eq:magnetic_codiff} are unchanged, while the electric Maxwell equation remains~\eqref{eq:elect_maxwell_first}. Using $\cE$, $\cB$, $p$ and $m_B$ defined above, we obtain
\begin{align}
    (\cD+2z)(\cD+\Delta)f
 &=-\frac{4\Lambda}{n}+\frac{2(n-1)\cE}{n}r^{-p}
 +\frac{2\cB}{n}r^{m_B},\label{eq:dyonic_general_tt}\\
 \alpha_\ell(\cD+\Delta)f
 &=R_{\hat\ell \hat\ell} r^{-2\kappa}-\frac{2\Lambda}{n}
 -\frac{\cE}{n}r^{-p}-v_\ell\cB r^{m_B},\label{eq:dyonic_general_ii}\\
 \sigma(\cD+\Delta)f
 &=\frac\lambda2\left(\cB r^{m_B}-\cE r^{-p}\right),\label{eq:dyonic_general_phi}\\
    zS&=Q+\sigma^2.\label{eq:dyonic_general_z}
\end{align}
The electric and magnetic contributions generally scale differently with radius. In the particularly simple branch
\begin{equation}
    \label{eq:dyonic_const_powers}
    p=m_B=0\quad\Longrightarrow\quad
    \lambda\sigma=-2S=2(\alpha_i+\alpha_j),
\end{equation}
both local electromagnetic energy densities have the same radial weight as the cosmological-constant term. Thus the fields support the asymptotic anisotropic geometry rather than acting as independent charge deformations of an otherwise fixed background. This is why their amplitudes are determined algebraically together with the scaling exponents. The running dilaton compensates for the radial dilution of the electric and magnetic fields.

If $\Lambda\neq0$ and $\lambda\sigma\neq0$, a nonzero spatial curvature then requires $\kappa=0$. Taking $(\cD+\Delta)f=\Delta f_0$, the algebraic equations become
\begin{align}
    f_0 &=\frac{-2\Lambda+(n-1)\cE+\cB}{z\Delta n},\label{eq:dyonic_const_1}\\
    R_\ell&=\alpha_\ell \Delta f_0 +\frac{2\Lambda+\cE}{n}+v_\ell\cB,\label{eq:dyonic_const_2}\\
    \sigma &=\frac\lambda{ 2 \Delta f_0}(\cB-\cE),\qquad
    zS=Q+\sigma^2,\label{eq:dyonic_const_3}\\
    f(r)&=f_0\left[1-\left(\frac{r_h}{r}\right)^\Delta\right].\label{eq:dyonic_const_f}
\end{align}
For a specified spatial Lie algebra and magnetic orientation, these equations constrain the exponents and matter amplitudes. The five-dimensional Nil solution and higher-dimensional examples are given in Sec.~\ref{sec:examples}.

The $\Lambda=0$ dyonic branches require a separate analysis, analogous to the neutral case in Sec.~\ref{sec:neutral_solutions}.

\paragraph{Remark on charged branches.}

The constant source configurations above use the electromagnetic fields to maintain the scaling background, so their amplitudes are not independent thermodynamic charges. One may instead seek solutions closer to a Reissner--Nordstr\"om black brane, in which a constant dilaton and balanced electric and magnetic fields introduce an additional, faster-decaying charge term in the blackening function, together with extremal horizons. Charged black holes of this type in Lifshitz spacetimes were constructed in \cite{Tarrio:2011de}, and on Thurston horizon geometries, including balanced dyonic configurations and their thermodynamics, in \cite{Bravo-Gaete:2017nkp,Arias:2017yqj,Arias:2018mqn}. Extending the present anisotropic scaling backgrounds with such independently variable charge deformations is left for future work.

\section{Examples of homogeneous black branes}
\label{sec:examples}

We now apply the reduced field equations to five-dimensional Bianchi geometries and higher-dimensional Ricci nilsolitons, beginning with the neutral solutions and then their matter-supported extensions.

\subsection{Neutral solutions in five dimensions}
\label{sec:examples_neutral}

We take $n=3$, with $\Lambda<0$, and use
\begin{equation}
    ds^2=-r^{2z}f(r)\dd t^2+\frac{\dd r^2}{r^2f(r)}
    +\sum_{i=1}^3b_i^2r^{2\alpha_i}(\theta^i)^2.
\end{equation}
The spatial one-forms and corresponding exponents are summarized below. Except where indicated, $f(r)=f_0[1-(r_h/r)^\Delta]$ with $f_0=-2\Lambda/(3z\Delta)$ and $\Delta=z+S$.

\paragraph{Bianchi II (Nil).}

The Nil coframe and scaling exponents are
\begin{equation}
    \theta^1=\dd x,\quad \theta^2=\dd y,\quad
    \theta^3=\dd w-x\dd y,\qquad
    (\alpha_1,\alpha_2,\alpha_3)=(1,1,2),
\end{equation}
so that $S=4$ and $Q=6$. We take $\Lambda<0$, $b_1=b_2=1$, and $b_3^2=-4\Lambda/9$.

The exponents satisfy the derivation condition \eqref{eq:kappa} with $\kappa=0$ and the mixed condition \eqref{eq:ri_constraint}, and \eqref{eq:vacuum_z} gives $z=Q/S=3/2$. The reference metric $h_b=(\theta^1)^2+(\theta^2)^2+b_3^2(\theta^3)^2$ has orthonormal Ricci eigenvalues
$R_1=R_2=-b_3^2/2$ and $R_3=b_3^2/2$, so the constraint \eqref{eq:neutral_ricci_constraint} fixes $b_3^2=-4\Lambda/9$. The blackening function follows from \eqref{eq:neutral_f} with $\Delta=z+S=11/2$.

\begin{align}\label{eq:neutral_nil_example}
    \dd s^2&=-r^3f(r)\dd t^2+\frac{\dd r^2}{r^2f(r)}
    +r^2[(\theta^1)^2+(\theta^2)^2]-\frac{4\Lambda}{9}r^4(\theta^3)^2,\\
    f(r)&=-\frac{8\Lambda}{99}\left[1-\left(\frac{r_h}{r}\right)^{11/2}\right].\nonumber
\end{align}

This is the Nil-horizon black brane already considered in \cite{Hassaine:2015ifa,Cadeau:2000tj}. The positive eigenvalue of $\Ric(h_b)$ in the third direction is compensated by the larger radial scaling exponent $\alpha_3$.

\paragraph{Solv (Bianchi VI$_{-1}$).}

The Solv coframe and scaling exponents are
\begin{equation}
    \theta^1=\e^w\dd x,\quad\theta^2=\e^{-w}\dd y,
    \quad\theta^3=\dd w,\qquad
    (\alpha_1,\alpha_2,\alpha_3)=(1,1,0),
\end{equation}
so that $S=Q=2$. We take $\Lambda<0$ and $b_1=b_2=1$.

As in the Nil example, \eqref{eq:kappa} and \eqref{eq:ri_constraint} hold with $\kappa=0$, and \eqref{eq:vacuum_z} gives $z=1$. The reference metric has orthonormal Ricci eigenvalues $R_1=R_2=0$ and $R_3=-2/b_3^2$, so \eqref{eq:neutral_ricci_constraint} fixes $b_3^2=-3/\Lambda$. The blackening function is \eqref{eq:neutral_f} with $\Delta=z+S=3$.

\begin{align}\label{eq:neutral_solv_example}
    \dd s^2&=-r^2f(r)\dd t^2+\frac{\dd r^2}{r^2f(r)}
    +r^2[(\theta^1)^2+(\theta^2)^2]-\frac3\Lambda(\theta^3)^2,\\
    f(r)&=-\frac{2\Lambda}{9}\left[1-\left(\frac{r_h}{r}\right)^3\right].\nonumber
\end{align}

Here $R_1=R_2=0$ and $R_3=2\Lambda/3$. This familiar solution is the special Bianchi VI$_{-1}$ case of the family below~\cite{Cadeau:2000tj,Hassaine:2015ifa}.

\paragraph{Bianchi VI$_h$.}

In our previous work~\cite{Amano:2026ayj} we considered the one-parameter family of coframes
\begin{equation}
    \label{eq:VIh_forms}
    \theta^1=\e^w\dd x,\qquad\theta^2=\e^{hw}\dd y,
    \qquad\theta^3=\dd w,
\end{equation}
with exponents
\begin{equation}
    (\alpha_1,\alpha_2,\alpha_3)=(-h,1,0),\qquad
    z=\frac{1+h^2}{1-h},\qquad
    \Delta=\frac{2(h^2-h+1)}{1-h},
\end{equation}
so that $S=1-h$ and $Q=1+h^2$. We take $h<1$, $\Lambda<0$, and $b_1=b_2=1$.

The mixed condition \eqref{eq:ri_constraint} reduces to $\alpha_1=-h\alpha_2$ for $\alpha_3=0$ and is satisfied by these exponents. The constraint \eqref{eq:neutral_ricci_constraint} applied to $h_b=(\theta^1)^2+(\theta^2)^2+b_3^2(\theta^3)^2$ fixes $b_3^2=-3(1+h^2)/(2\Lambda)$ and is then satisfied for every $h<1$. The blackening function is \eqref{eq:neutral_f} with $f_0=-\Lambda(1-h)^2/[3(1+h^2)(h^2-h+1)]$.

\begin{align}\label{eq:neutral_vih}
    \dd s^2={}&-r^{2z}f(r)\dd t^2+\frac{\dd r^2}{r^2f(r)}
    +r^{-2h}(\theta^1)^2+r^2(\theta^2)^2
    -\frac{3(1+h^2)}{2\Lambda}(\theta^3)^2,\\
    f(r)={}&-\frac{\Lambda(1-h)^2}{3(1+h^2)(h^2-h+1)}
    \left[1-\left(\frac{r_h}{r}\right)^\Delta\right].\nonumber
\end{align}

The limiting value $h=-1$ gives~\eqref{eq:neutral_solv_example}. The $h=0$ case has $Q=S^2$, where the radial root is repeated, and must be interpreted with the corresponding equations; it still admits the displayed black brane. The point $h=1$ is a separate Bianchi V branch, for which the radial ansatz must be modified. These cases and the associated Ricci-flat, hyperscaling-violating family were derived in Ref.~\cite{Amano:2026ayj}.

\subsection{Electric five-dimensional solutions}

We next turn to EMD gravity. For concreteness, we use the constant source electric branch~\eqref{eq:elect_const_source}. In each case the dilaton takes the form $\phi=\phi_0+\sigma\log r$ and the gauge field is given by~\eqref{eq:elect_maxwell_first}. The field strength therefore shares the spatial homogeneity of the metric.

\paragraph{Bianchi I.}

For a flat spatial metric we take $\theta^i=\dd x^i$ with $(\alpha_1,\alpha_2,\alpha_3)=(1,1,1)$, so that $S=Q=3$. We take $\Lambda<0$, $\lambda\neq0$, and $b_i=1$.

The constant source branch \eqref{eq:elect_const_source} gives $\lambda\sigma=-6$, and \eqref{eq:emd_zS} gives $z=1+12/\lambda^2$. The amplitudes \eqref{eq:elect_amplitude} give $\cE=-8\Lambda/(\lambda^2+4)$, and \eqref{eq:elect_blackening} gives $\Delta=4+12/\lambda^2$. Since the spatial geometry is flat, $R_{\hat i \hat i}=0$, the spatial equations \eqref{eq:elect_algebraic} are satisfied identically by this data.

\begin{align}\label{eq:elect_flat_solution}
    \dd s^2&=-r^{2z}f(r)\dd t^2+\frac{\dd r^2}{r^2f(r)}
    +r^2(\dd x^2+\dd y^2+\dd w^2),\\
    f(r)&=-\frac{\Lambda\lambda^4}{6(\lambda^2+4)(\lambda^2+3)}
    \left[1-\left(\frac{r_h}{r}\right)^\Delta\right],
    \qquad
    \phi=\phi_0-\frac6\lambda\log r,\nonumber\\
    F&=q\,\e^{-\lambda\phi_0}\,r^{\,3+12/\lambda^2}\dd r\wedge\dd t,
    \qquad
    q^2\e^{-\lambda\phi_0}=\cE=-\frac{8\Lambda}{\lambda^2+4}.\nonumber
\end{align}

This is a homogeneous Lifshitz-type solution supported by a running dilaton and an electric field. It is useful as a check on the normalization of our equations.

\paragraph{Bianchi II.}

We use the Nil coframe above with $(\alpha_1,\alpha_2,\alpha_3)=(1,1,2)$, so that $S=4$ and $Q=6$. We take $\Lambda<0$, $\lambda\neq0$, and $b_1=b_2=1$.

The constant source branch \eqref{eq:elect_const_source} gives $\lambda\sigma=-8$, and \eqref{eq:emd_zS} gives $z=\frac32+\frac{16}{\lambda^2}$. The amplitudes \eqref{eq:elect_amplitude} give $\cE=-\frac{64\Lambda}{32+9\lambda^2}$ and $\Delta=\frac{11}{2}+\frac{16}{\lambda^2}$. The Nil Ricci data $R_1=R_2=-b_3^2/2$ and $R_3=b_3^2/2$ in the spatial equations \eqref{eq:elect_algebraic} fix $b_3^2=\lambda^2\cE/16$.

\begin{align}\label{eq:elect_nil_solution}
    \dd s^2&=-r^{2z}f(r)\dd t^2+\frac{\dd r^2}{r^2f(r)}
    +r^2\left[(\theta^1)^2+(\theta^2)^2\right]+b_3^2r^4(\theta^3)^2,\\
    f(r)&=-\frac{8\Lambda\lambda^4}{(9\lambda^2+32)(11\lambda^2+32)}
    \left[1-\left(\frac{r_h}{r}\right)^\Delta\right],
    \qquad
    \phi=\phi_0-\frac8\lambda\log r,\nonumber\\
    F&=q\,\e^{-\lambda\phi_0}\,r^{\,9/2+16/\lambda^2}\dd r\wedge\dd t,
    \qquad
    q^2\e^{-\lambda\phi_0}=\cE=-\frac{64\Lambda}{9\lambda^2+32},\nonumber
\end{align}
with $b_3^2=-4\Lambda\lambda^2/(9\lambda^2+32)$.

The scalar and electric field modify $z$ while preserving the underlying Nil spatial geometry. In the limit $\lambda\to\infty$ at fixed $\Lambda$ the neutral solution \eqref{eq:neutral_nil_example} is recovered.

\paragraph{Bianchi VI$_h$.}

We may extend the neutral family \eqref{eq:neutral_vih} with the same coframe \eqref{eq:VIh_forms} and exponents $(-h,1,0)$, so that $S=1-h$ and $Q=1+h^2$. We take $h<1$, $\Lambda<0$, $\lambda\neq0$, and $b_1=b_2=1$.

The constant source branch \eqref{eq:elect_const_source} gives $\lambda\sigma=-2(1-h)$, and \eqref{eq:emd_zS} gives $z=\frac{1+h^2}{1-h}+\frac{4(1-h)}{\lambda^2}$. The amplitudes \eqref{eq:elect_amplitude} give
\begin{equation}\label{eq:elect_vi_parameters}
    \cE=-\frac{8\Lambda(1-h)^2}{4(1-h)^2+3\lambda^2(1+h^2)},\qquad
    b_3^2=-\frac{4(1-h)^2+3\lambda^2(1+h^2)}{2\Lambda\lambda^2},
\end{equation}
where $b_3^2$ follows from the spatial equations \eqref{eq:elect_algebraic}. The blackening function is \eqref{eq:elect_blackening} with $\Delta=z+1-h$.

\begin{align}\label{eq:elect_vih_solution}
    \dd s^2&=-r^{2z}f(r)\dd t^2+\frac{\dd r^2}{r^2f(r)}
    +r^{-2h}(\theta^1)^2+r^2(\theta^2)^2+b_3^2(\theta^3)^2,\\
    f(r)&=f_0\left[1-\left(\frac{r_h}{r}\right)^\Delta\right],
    \qquad\phi=\phi_0-\frac{2(1-h)}\lambda\log r,\nonumber\\
    F&=q\,\e^{-\lambda\phi_0}\,r^{\Delta-1}\dd r\wedge\dd t,
    \qquad
    q^2\e^{-\lambda\phi_0}=\cE,\nonumber\\
    f_0&=-\frac{2\Lambda(1-h)^2\lambda^4}
    {\left[3\lambda^2(1+h^2)+4(1-h)^2\right]
    \left[\lambda^2(2-2h+2h^2)+4(1-h)^2\right]}.\nonumber
\end{align}

The radial electric field modifies the time exponent and the curvature scale, while the continuous Bianchi parameter $h$ is retained. In the limit of vanishing matter this is not necessarily a continuous limit at fixed $\lambda$, because the relation $\lambda\sigma=-2S$ selects a specific charged branch.

\subsection{Magnetic and dyonic five-dimensional solutions}

\paragraph{Magnetic Bianchi I.}

We consider a five-dimensional spacetime with flat spatial
sections, $\theta^i=\dd x^i$, and a magnetic field in the
$(12)$-plane,
\begin{equation}
    F=B\,\dd x\wedge\dd y.
\end{equation}
Although the spatial geometry is flat, the magnetic field
introduces anisotropy between the magnetic plane and the
remaining direction. We take $\lambda\neq0$ and
$\Lambda<0$; the radial diffeomorphism
\eqref{eq:radial_rescale} fixes the exponent scale to
$a=1$.

The magnetic Bianchi identity and Maxwell equation
\eqref{eq:magnetic_closure}--\eqref{eq:magnetic_codiff}
hold trivially since $\dd\theta^i=0$. The conditions
$m_B=0$ and $\kappa=0$ give $\lambda\sigma=4$, and
the spatial equations give $\alpha_3=z$. The constraint
$zS=Q+\sigma^2$ then yields $z=1+8/\lambda^2$. Thus the
exponents are $(\alpha_1,\alpha_2,\alpha_3)=(1,1,z)$ and
$\Delta=2(1+z)$. The algebraic system
\eqref{eq:mag_algebraic} with $R_{\hat i \hat i}=0$ then gives
$\cB=-2\Lambda(z-1)/(1+2z)=-16\Lambda/(3\lambda^2+16)$
and $f_0=-2\Lambda/[(1+2z)\Delta]$.

\begin{align}
    \dd s^2={}&-r^{2z}f(r)\dd t^2
    +\frac{\dd r^2}{r^2f(r)}
    +r^{2}(\dd x^2+\dd y^2)
    +r^{2z}\dd w^2,\\
    f(r)&={}-\frac{2\Lambda}{(1+2z)\Delta}
    \left[1-\left(\frac{r_h}{r}\right)^\Delta\right],
    \qquad
    \phi=\phi_0+\frac4\lambda\log r,\nonumber\\
    F&=B\,\dd x\wedge\dd y,
    \qquad
    B^2\e^{\lambda\phi_0}=\cB=-\frac{16\Lambda}{3\lambda^2+16}.\nonumber
\end{align}

For $\Lambda<0$, the magnetic amplitude and blackening
function are positive outside the horizon. The magnetic
field supports different radial scaling in the magnetic
plane and the remaining spatial direction, despite the
vanishing intrinsic spatial curvature. The solution
approaches the neutral planar AdS black brane as
$\lambda^2\rightarrow\infty$: the exponents become isotropic,
and the scalar gradient and electromagnetic stress vanish.

\paragraph{Magnetic Solv.}

Consider the Solv coframe above with $(\alpha_1,\alpha_2,\alpha_3)=(1,1,0)$ and a magnetic field in the $(12)$-plane,
\begin{equation}
    F=B\theta^1\wedge\theta^2.
\end{equation}
We take $\Lambda<0$, $\lambda\neq0$, and $b_1=b_2=1$.

Since $\dd(\theta^1\wedge\theta^2)=0$ and $\dd\theta^3=0$, both the Bianchi identity \eqref{eq:magnetic_closure} and the Maxwell equation \eqref{eq:magnetic_codiff} are satisfied. The conditions $m_B=0$ and $\kappa=0$ give $\lambda\sigma=4$, and $zS=Q+\sigma^2$ gives $z=1+8/\lambda^2$; the radial diffeomorphism \eqref{eq:radial_rescale} fixes the exponent scale to $a=1$. The algebraic system \eqref{eq:mag_algebraic}, with the Solv Ricci data $R_1=R_2=0$ and $R_3=-2/b_3^2$, gives
\begin{equation}
    \label{eq:mag_solv_par}
    \sigma=\frac4\lambda,\qquad
    z=1+\frac8{\lambda^2},\qquad
    \cB=-\frac{2\Lambda(z-1)}{1+2z},\qquad
    b_3^2=-\frac{1+2z}{z\Lambda}.
\end{equation}

\begin{align}\label{eq:mag_solv_fields}
    \dd s^2&=-r^{2z}f(r)\dd t^2+\frac{\dd r^2}{r^2f(r)}
    +r^{2}\left[(\theta^1)^2+(\theta^2)^2\right]+b_3^2(\theta^3)^2,\\
    f(r)&=-\frac{2\Lambda}{(1+2z)(z+2)}
    \left[1-\left(\frac{r_h}{r}\right)^{z+2}\right],
    \qquad
    \phi=\phi_0+\frac4\lambda\log r,\nonumber\\
    F&=B\,\theta^1\wedge\theta^2,
    \qquad
    B^2\e^{\lambda\phi_0}=\cB.\nonumber
\end{align}

For $\Lambda<0$, the solution has a simple horizon and positive magnetic amplitude. The different signs of the electric and magnetic contributions to the scalar equation explain the opposite sign of $\sigma$ compared with the electric examples.

\paragraph{Dyonic Nil.}

The Nil group also admits a configuration with both electric and magnetic fields. We use the Nil coframe with $(\alpha_1,\alpha_2,\alpha_3)=(1,-3/4,1/4)$, so that $S=1/2$ and $Q=13/8$, and choose the magnetic plane $(23)$. The field strength is
\begin{equation}
    F=E(r)\dd r\wedge\dd t+B_{23}\theta^2\wedge\theta^3.
\end{equation}
We take $\lambda\neq0$ and $\Lambda<0$; the radial diffeomorphism \eqref{eq:radial_rescale} fixes the exponent scale to $a=1$.

The magnetic form is closed, and its spatial dual is proportional to the closed form $\theta^1$, so the constraints \eqref{eq:magnetic_closure}--\eqref{eq:magnetic_codiff} hold. The constant source conditions \eqref{eq:dyonic_const_powers} give $\lambda\sigma=-1$, and \eqref{eq:dyonic_general_z} gives $z=\frac{13}{4}+\frac{2}{\lambda^2}$ and $\Delta=\frac{15}{4}+\frac{2}{\lambda^2}$. The algebraic equations \eqref{eq:dyonic_const_1}--\eqref{eq:dyonic_const_3}, with the Nil Ricci data $R_1=R_2=-b_3^2/2$ and $R_3=b_3^2/2$, have the solution
\begin{align}\label{eq:dyonic_nil_par}
    \sigma&=-1/\lambda,
          &z&=\frac{13}{4}+\frac{2}{\lambda^2},\\
    \Delta&=\frac{15}{4}+\frac{2}{\lambda^2},
          &\cB&=-\frac{7\Lambda\lambda^2}{9\lambda^2+4},\\
    \cE&=-\frac{\Lambda(7\lambda^2+8)}{9\lambda^2+4},
       &b_3^2&=-\frac{4\Lambda\lambda^2}{9\lambda^2+4}
       b_1^2b_2^2.\nonumber
\end{align}
\begin{align}\label{eq:dyonic_nil_fields}
    \dd s^2={}&-r^{2z}f(r)\dd t^2+\frac{\dd r^2}{r^2f(r)}
    +b_1^2r^{2}(\theta^1)^2+b_2^2r^{-3/2}(\theta^2)^2
    +b_3^2r^{1/2}(\theta^3)^2,\\
    f(r)={}&-\frac{4\Lambda\lambda^2}
    {(9\lambda^2+4)\Delta}
    \left[1-\left(\frac{r_h}{r}\right)^\Delta\right],
    \qquad
    \phi=\phi_0-\frac{1}{\lambda}\log r,\nonumber\\
    F={}&q\e^{-\lambda\phi_0}r^{z+S-1}\dd r\wedge\dd t
    +B_{23}\theta^2\wedge\theta^3,\nonumber
\end{align}
where
\begin{equation}
    q^2\e^{-\lambda\phi_0}
    =-\frac{\Lambda(7\lambda^2+8)}{9\lambda^2+4},
    \qquad
    \frac{B_{23}^2\e^{\lambda\phi_0}}{b_2^2b_3^2}
    =-\frac{7\Lambda\lambda^2}{9\lambda^2+4}.
\end{equation}

For $\Lambda<0$, both electric and magnetic amplitudes are positive and the blackening function has a single horizon. The running dilaton supports a homogeneous anisotropic dyonic configuration in which both sources enter the same constant part of the reduced Einstein equations.

\subsection{Higher-dimensional families}
\label{sec:examples_higher}

\paragraph{Dyonic Bianchi I in general dimensions.}

We now consider a flat spatial geometry in $D=n+2$
dimensions, with $n\geq3$, and introduce electric and
magnetic fields,
\begin{equation}
    F=E(r)\dd r\wedge\dd t+B\,\dd x^1\wedge\dd x^2.
\end{equation}
We take $\lambda\neq0$ and $\Lambda<0$; the radial diffeomorphism \eqref{eq:radial_rescale} fixes the exponent scale to $a=1$.

For the constant source branch, the conditions
$p=m_B=0$ in \eqref{eq:dyonic_const_powers} require
$\lambda\sigma=-2S=2(\alpha_1+\alpha_2)$. Since the
intrinsic spatial Ricci tensor vanishes, the spatial
Einstein equations require equal scaling exponents within
the magnetic plane and within its orthogonal complement.
The algebraic equations
\eqref{eq:dyonic_const_1}--\eqref{eq:dyonic_const_3}
then give the exponents and amplitudes
\begin{align}
    \alpha_1=\alpha_2&=-1,
                     &\alpha_A&=\frac{4}{n-2},
                     \qquad A=3,\ldots,n,\\
    S&=2,
     &\sigma&=-\frac{4}{\lambda},\\
    z&=1+\frac{8}{n-2}
    +\frac{8}{\lambda^2},
     &\Delta&=z+2.
\end{align}
The remaining algebraic equations give
\begin{align}
    \cB&=\left(1+\frac{4}{n-2}\right)\Delta f_0,\\
    \cE&=\left(z-\frac{4}{n-2}\right)\Delta f_0.
\end{align}

\begin{align}
    \dd s^2={}&-r^{2z}f(r)\dd t^2
    +\frac{\dd r^2}{r^2f(r)}
    +r^{-2}\left[(\dd x^1)^2+(\dd x^2)^2\right]
    +r^{\frac{8}{n-2}}
    \sum_{A=3}^n(\dd x^A)^2,\\
    f(r)&={}-\frac{2\Lambda}{(z+3)(z+2)}
    \left[1-\left(\frac{r_h}{r}\right)^\Delta\right],
    \qquad
    \phi=\phi_0-\frac4\lambda\log r,\nonumber\\
    F&={}q\e^{-\lambda\phi_0}r^{z+1}
    \dd r\wedge\dd t
    +B\,\dd x^1\wedge\dd x^2,
    \qquad
    q^2\e^{-\lambda\phi_0}=\cE,\quad
    B^2\e^{\lambda\phi_0}=\cB.\nonumber
\end{align}

For $\Lambda<0$, both $\cE$ and $\cB$ are positive,
and $f(r)$ has a simple horizon at $r=r_h$.
In contrast to the neutral and purely electric flat
solutions, the magnetic directions have negative
scaling exponents, while the remaining directions
have positive exponents. Both electromagnetic sources
contribute to the constant part of the Einstein
equations, rather than introducing an independent
charge-dependent radial mode in the blackening
function.

\paragraph{Neutral Ricci nilsolitons.}

The neutral construction of Sec.~\ref{sec:neutral} extends to arbitrary dimensions, as first considered in Ref.~\cite{Hervik:2007zz}. The input is an $n$-dimensional nilpotent Lie algebra with a left-invariant metric $h_b$ satisfying
\begin{equation}
    \label{eq:nilsoliton_input}
    \Ric(h_b)=-c\,\mathbf1+\mathsf D,\qquad c>0,
    \qquad \mathsf D(X_i)=d_i X_i,
\end{equation}
where $\mathsf D$ is a symmetric derivation. We take $\Lambda<0$ and a free time exponent $z>0$, which plays the role of the overall exponent scale $a$ and can likewise be set to unity by \eqref{eq:radial_rescale}.

Rescale $h_b$ so that $c=2(-\Lambda)/n$. The Ricci constraint \eqref{eq:neutral_ricci_constraint} is then solved by the soliton data
\begin{equation}
    \label{eq:nilsoliton_neutral_exp}
    \alpha_i=\frac{z}{c}d_i,\qquad
    S=\frac{z}{c}\Tr\mathsf D,
    \qquad
    f_0=\frac{c}{z(z+S)}.
\end{equation}
For a nonflat Ricci nilsoliton, the identity $\Tr\mathsf D^2=c\Tr\mathsf D$ gives $Q=zS$, so \eqref{eq:vacuum_z} holds.

\begin{equation}
    \label{eq:nilsoliton_neutral_metric}
    \dd s^2=-r^{2z}f(r)\dd t^2+\frac{\dd r^2}{r^2f(r)}
    +\sum_i b_i^2r^{2zd_i/c}(\theta^i)^2,
    \qquad f(r)=f_0\left[1-\left(\frac{r_h}{r}\right)^{z+S}\right].
\end{equation}

The metric solves the neutral Einstein equations, subject to the mixed constraints. This is the Ricci nilsoliton black hole construction of Ref.~\cite{Hervik:2007zz}.

\paragraph{Dyonic extension of a Ricci nilsoliton.}

We add two flat spatial directions to an $m$-dimensional nonflat Ricci nilsoliton. Let
\begin{equation}
    \Ric(h_0)=-c\,\mathbf1+\mathsf D,
    \qquad
    \mathsf D(X_i)=d_i X_i,
    \qquad
    \sum_i d_i^2=c\sum_i d_i.
\end{equation}
We use $\mathfrak g=\mathfrak n_m\oplus\mathbb R^2$ with coordinates $(x,y)$ along the additional flat directions. The total spatial dimension is $n=m+2$ and the spacetime dimension is $D=m+4$. We take $\lambda\neq0$ and $\Lambda<0$, and a magnetic field in the additional flat plane, $F=E(r)\dd r\wedge\dd t+B\,\dd x\wedge\dd y$.

The radial diffeomorphism \eqref{eq:radial_rescale} fixes the exponent scale to $a=1$. The dyonic constant source conditions \eqref{eq:dyonic_const_powers} are met by the soliton data below, and the algebraic equations \eqref{eq:dyonic_const_1}--\eqref{eq:dyonic_const_3} give
\begin{align}\label{eq:nil_dyonic_param_general}
    \alpha_i&=d_i\quad(i=1,\ldots,m),
            &\alpha_x=\alpha_y&=-\frac{1}{4}\sum_i d_i,\\
    S&=\frac{1}{2}\sum_i d_i,
     &\sigma&=-\frac{1}{\lambda}\sum_i d_i,\\
    z&=2c+
    \left(\frac14+\frac2{\lambda^2}\right)\sum_i d_i,
     &\Delta&=z+S,\\
    L^2&=\frac{-nc-\frac2{\lambda^2}\sum_i d_i}{2\Lambda},
       &\cB&=\frac{c+\frac14\sum_i d_i}{L^2},\\
    \cE&=\frac{c+
    \left(\frac14+\frac2{\lambda^2}\right)\sum_i d_i}{L^2}.
    \nonumber
    \end{align}

    We take a representative nilsoliton metric $h_0=\sum_i\bar b_i^2(\theta^i)^2$ and obtain
    \begin{align}\label{eq:nil_dyonic_metric_general}
        \dd s^2={}&-r^{2z}f(r)\dd t^2+\frac{\dd r^2}{r^2f(r)}
        +L^2\sum_{i=1}^m\bar b_i^2r^{2d_i}(\theta^i)^2
        +L^2r^{-\frac12\sum_i d_i}(\dd x^2+\dd y^2),\\
        f(r)&=\frac{1}{L^2\Delta}
        \left[1-\left(\frac{r_h}{r}\right)^\Delta\right],
        \qquad
        \phi=\phi_0-\frac{1}{\lambda}\sum_i d_i\log r,
        \nonumber\\
        F&=q\e^{-\lambda\phi_0}r^{z+S-1}\dd r\wedge\dd t
        +B\dd x\wedge\dd y,\nonumber\\
        q^2\e^{-\lambda\phi_0}&=\cE,
        \qquad
        \frac{B^2\e^{\lambda\phi_0}}{L^4}=\cB.
        \nonumber
    \end{align}
    The two matter source powers are constant, $p=m_B=0$. The construction also obeys the scalar constraint and the temporal equation because $\sum_i d_i^2=c\sum_i d_i$. The radial momentum constraint follows from the divergence-free property of the soliton derivation. The magnetic form is closed, while its dual is proportional to the volume form on the nilpotent factor; hence the Maxwell equation also holds.
    The additional magnetic plane extends the neutral Ricci nilsoliton construction to a family of higher-dimensional dyonic solutions.

    As one example, take the three-dimensional Nil soliton with $c=3/2$ and $(d_1,d_2,d_3)=(1,1,2)$, so that $n=5$ and $D=7$. In this case,
    \begin{equation}
        \label{eq:seven_D_example}
        \alpha_x=\alpha_y=-1,\quad
        S=2,\quad \sigma=-\frac{4}{\lambda},\quad
        z=4+\frac8{\lambda^2},\quad
        L^2=-\frac{15+16/\lambda^2}{4\Lambda},
    \end{equation}
    with $\Delta=6+8/\lambda^2$, $\cB=5/(2L^2)$, and $\cE=(5/2+8/\lambda^2)/L^2$, so that
    \begin{align}
        \dd s^2={}&-r^{2z}f(r)\dd t^2+\frac{\dd r^2}{r^2f(r)}
        +L^2\left[\bar b_1^2r^{2}(\theta^1)^2+\bar b_2^2r^{2}(\theta^2)^2
        +\bar b_3^2r^{4}(\theta^3)^2+r^{-2}(\dd x^2+\dd y^2)\right],\\
        f(r)&={}\frac{1}{L^2\Delta}
        \left[1-\left(\frac{r_h}{r}\right)^\Delta\right],
        \qquad
        \phi=\phi_0-\frac4\lambda\log r,\nonumber\\
        F&={}q\e^{-\lambda\phi_0}r^{z+1}\dd r\wedge\dd t
        +B\,\dd x\wedge\dd y,
        \qquad
        q^2\e^{-\lambda\phi_0}=\cE,\quad
        \frac{B^2\e^{\lambda\phi_0}}{L^4}=\cB.\nonumber
    \end{align}
    The transverse magnetic directions shrink with increasing $r$, whereas the Nil directions grow. This illustrates how additional spatial dimensions can produce strongly anisotropic exact black brane geometries.

    \section{Thermodynamics}
    \label{sec:thermo}

    We next examine the physical properties of solutions with the blackening function \eqref{eq:intro_blackening}. These include the $\kappa=0$, $\Lambda<0$ neutral branch of Sec.~\ref{sec:neutral} and the constant-source electric, magnetic, and dyonic solutions in Sec.~\ref{sec:matter}.

    \subsection{Temperature and entropy}

    Let
    \begin{align}\label{eq:thermo_family}
        \dd s^2&=-r^{2z}f(r)\dd t^2+\frac{\dd r^2}{r^2f(r)}
        +\sum_i b_i^2r^{2\alpha_i}(\theta^i)^2,\\
        f(r)&=f_0-\frac\mu{r^\Delta}
        =f_0\left[1-\left(\frac{r_h}{r}\right)^\Delta\right],
        \qquad\mu=f_0r_h^\Delta.\nonumber
    \end{align}
    We take $f_0>0$, $z>0$, $S>0$, and $\Delta=z+S>0$. The horizon is at $r=r_h>0$ and the exterior region is $r>r_h$. The Euclidean regularity condition gives the Hawking temperature
    \begin{equation}
        \label{eq:th_temp}
        T=\frac{r_h^{z+1}f'(r_h)}{4\pi} =\frac{f_0\Delta}{4\pi}r_h^z.
    \end{equation}
    Let $\Sigma$ be a horizon spatial section. Its area is proportional to the volume of the reference left-invariant spatial metric. Define
    \begin{equation}
        \cV_b:=\int_\Sigma b_1\cdots b_n\,
        \theta^1\wedge\cdots\wedge\theta^n.
    \end{equation}
    If the spatial geometry has no compact quotient we instead use a fixed regulated cell and interpret the results as densities. The Bekenstein--Hawking entropy is
    \begin{equation}
        \label{eq:th_entropy}
        \mathcal S_{\mathrm{BH}}=2\pi\cV_br_h^S.
    \end{equation}
    The relation between entropy and temperature is therefore
    \begin{equation}
        \label{eq:entropy_scaling}
        \mathcal S_{\mathrm{BH}}=2\pi\cV_b
        \left(\frac{4\pi T}{f_0\Delta}\right)^{S/z}.
    \end{equation}
    We see that the scaling exponent $S/z$ controls the entropy. In the isotropic AdS case, $\alpha_i=z=1$ and $S=n$, so the familiar result $\mathcal S_{\mathrm{BH}}\sim T^n$ is recovered. For an anisotropic geometry, $S$ is the sum of the individual spatial scaling exponents, rather than the number of spatial dimensions. This temperature scaling is analogous to that found in Lifshitz black holes \cite{Liu:2014dva}, and the dilatonic Lifshitz families of \cite{Romero-Figueroa:2024sac} exhibit the corresponding extended thermodynamics.

    \subsection{Euclidean on-shell action and free energy at fixed flux}
    \label{sec:thermo_action}

    The fixed-flux free energy follows from the Euclidean on-shell action. Because the geometries are generally not asymptotically
    locally AdS, the action must be supplemented by boundary terms appropriate to
    the chosen asymptotic data. The radial Hamilton--Jacobi approach to these
    boundary terms was developed in Ref.~\cite{Papadimitriou:2005ii,Papadimitriou:2011qb, Papadimitriou:2010as} and
    extended to nonrelativistic asymptotics in
    Ref.~\cite{Chemissany:2014xsa}. Here we perform an explicit
    background subtraction within the restricted family~\eqref{eq:thermo_family},
    with all supporting sources and fluxes held fixed. This subtraction determines the relative
    on-shell action for variations within the fixed-source family.

    Under $t=-i\tau$, the electric potential is continued as
    $A_\tau=-iA_t$, and the Euclidean time period is $\beta:=1/T$.
    At a large radial cutoff $r=R$, let $\gamma_{ab}$ be the induced metric,
    $\gamma:=\det\gamma_{ab}$, and $K$ the trace of the extrinsic curvature
    with the outward normal. We take
    \begin{equation}
        I_E:=I_{\mathrm{bulk}}^E+I_{\mathrm{GHY}}^E+I_N^E+I_{\mathrm{ct}},
        \qquad
        I_{\mathrm{GHY}}^E:=-\int_{r=R}\dd^{n+1}x\sqrt\gamma\,K,
        \label{eq:thermo_ie_def}
    \end{equation}
    Let $n_\mu$ be the outward unit normal and $A_E$ the Euclidean gauge
    potential, with field strength $F_E:=\dd A_E$ and boundary components $A_a$.
    Fixing the radial electric flux, rather than the boundary potential, amounts
    to adding the Euclidean Legendre term
    \begin{equation}
        I_N^E:=-\int_{r=R}\dd^{n+1}x\sqrt\gamma\,
        e^{\lambda\phi}n_\mu F_E^{\mu a}A_a.
        \label{eq:thermo_neumann}
    \end{equation}
    This distinction between electric ensembles is familiar from black-hole
    thermodynamics~\cite{Hawking:1995ap}. We set $I_N^E=0$ in the neutral and
    purely magnetic cases.

    Taking the trace of~\eqref{eq:emd_einstein}, the expression in square brackets
    in the bulk action~\eqref{eq:emd_action} simplifies on shell to
    \begin{equation}
        \left.\left[R-2\Lambda-(\partial\phi)^2
            -\frac12 e^{\lambda\phi}F^2\right]\right|_{\mathrm{os}}
            =\frac{4\Lambda}{n}-\frac1n e^{\lambda\phi}F^2.
            \label{eq:thermo_trace_on_shell}
        \end{equation}
        We now specialize to the case where $\kappa=0$ and the local electric and
        magnetic energy densities are constant in $r$ (including the cases in which
        one or both fields are absent). With the conventions of Sec.~\ref{sec:matter},
        $e^{\lambda\phi}F^2=2(\cB-\cE)$ after the Euclidean continuation, and the
        temporal Einstein equation yields
        \begin{equation}
            \frac{4\Lambda+2\cE-2\cB}{n}
            =2(\cE-z\Delta f_0).
            \label{eq:thermo_lagr_on_shell}
        \end{equation}
        For the ansatz~\eqref{eq:thermo_family}, the volume element is independent of
        $f$ and the extrinsic curvature at $r=R$ is
        \begin{equation}
            \sqrt{g_E}\,\dd^{n+2}x=\dd\tau\,\dd r\,
            r^{\Delta-1}\,\dd\operatorname{vol}_{h_b},\qquad
            K=R\sqrt{f(R)}\left[\Delta+\frac{Rf'(R)}{2f(R)}\right].
            \label{eq:thermo_volume_k}
        \end{equation}
        The Euclidean bulk and Gibbons--Hawking terms therefore evaluate to
        \begin{align}
            \frac{I_{\mathrm{bulk}}^E}{\beta\cV_b}
 &=\left(zf_0-\frac{\cE}{\Delta}\right)
 (R^\Delta-r_h^\Delta),\\
 \frac{I_{\mathrm{GHY}}^E}{\beta\cV_b}
 &=-\left(\Delta f_0R^\Delta-\frac{\Delta\mu}{2}\right),
 \qquad \mu=f_0r_h^\Delta.
 \label{eq:thermo_bulk_ghy}
        \end{align}
        For $p=0$ the horizon-regular Lorentzian potential is
        $A_t=q e^{-\lambda\phi_0}(r^\Delta-r_h^\Delta)/\Delta$.
        With $\cE=q^2e^{-\lambda\phi_0}$, \eqref{eq:thermo_neumann}
        gives
        \begin{equation}
            \frac{I_N^E}{\beta\cV_b}
            =\frac{\cE}{\Delta}
            (R^\Delta-r_h^\Delta).
            \label{eq:thermo_neumann_eval}
        \end{equation}
        The $\cE$-dependent terms in the bulk action and the electric Legendre term
        cancel. Hence, prior to counterterms, the regulated canonical action is
        \begin{equation}
            \frac{I_{E,\mathrm{reg}}^{(N)}}{\beta\cV_b}
            =\frac12
            \left[-2Sf_0R^\Delta+(S-z)\mu\right].
            \label{eq:thermo_regulated_action}
        \end{equation}

        One way to obtain the finite \emph{relative} action is to subtract the
        $\mu=0$ scaling geometry with precisely the same $\Lambda$, $b_i$, $\alpha_i$,
        $\phi_0$, electric flux, and magnetic flux.
        The reference fixes the zero of the relative action within the family
        with fixed asymptotic sources. To match the proper length of the
        Euclidean time circle at the cutoff, its period is
        \begin{equation}
            \beta_0=\beta\sqrt{\frac{f(R)}{f_0}}
            =\beta\left[1-\frac{\mu}{2f_0R^\Delta}
            +O(R^{-2\Delta})\right].
            \label{eq:thermo_period_match}
        \end{equation}
        Since the reference action is
        $I_{E,0}^{(N)}=-\beta_0\cV_b Sf_0R^\Delta$,
        we obtain
        \begin{align}
            \lim_{R\to\infty}
            \frac{I_{E,\mathrm{reg}}^{(N)}[\mu]
            -I_{E,\mathrm{reg}}^{(N)}[0]}{\beta\cV_b}
 &=\frac12
 \lim_{R\to\infty}
 \left[-2Sf_0R^\Delta\left(1-\sqrt{1-\frac{\mu}{f_0R^\Delta}}\right)
 +(S-z)\mu\right]
 \nonumber\\
 &=-\frac{z\mu}{2}.
 \label{eq:thermo_action_difference}
        \end{align}
        Alternatively, the \emph{on-shell} divergence of this family is removed by
        \begin{equation}
            I_{\mathrm{ct}}^{\mathrm{fam}}
            :=S\sqrt{f_0}\int_{r=R}\dd^{n+1}x\sqrt\gamma,
            \label{eq:thermo_ct_family}
        \end{equation}
        which gives the same finite result for the black brane. In planar AdS,
        $S=n$ and $f_0=L^{-2}$, and this reduces to the usual leading
        $n/L$ counterterm. In a general anisotropic background, however,
        \eqref{eq:thermo_ct_family} is only a counterterm evaluated on the
        restricted solution family. It need not provide finite one-point functions or
        a well-posed variational problem under arbitrary variations of the anisotropic
        metric and matter sources. A complete Hamilton--Jacobi counterterm functional \cite{Papadimitriou:2011qb}
        may include intrinsic-curvature and matter invariants at the same asymptotic
        order.

        We may therefore identify the fixed-flux free energy relative to the
        scaling background as
        \begin{equation}
            \label{eq:th_F}
            \mathcal F_N-\mathcal F_0
            =\frac{I_{E,\mathrm{ren}}^{(N)}}\beta
            =-\frac{\cV_b z}{2}\mu
            =-\frac{\cV_b z f_0}{2}r_h^\Delta.
        \end{equation}
        Combining this result with the independently obtained horizon entropy and
        temperature, the associated thermodynamic energy difference is
        \begin{equation}
            \label{eq:th_M}
            M-M_0=(\mathcal F_N-\mathcal F_0)
            +T\mathcal S_{\mathrm{BH}}
            =\frac{\cV_b S}{2}\mu,
            \qquad
            \Delta(M-M_0)=S T\mathcal S_{\mathrm{BH}}.
        \end{equation}
        The resulting energy agrees with the restricted first law and the
        Hamiltonian variation in Appendix~\ref{app:mass}. At fixed supporting sources, the heat capacity is
        \begin{equation}
            \label{eq:th_C}
            C_{\mathrm{fixed}}:=T
            \left(\frac{\pd\mathcal S_{\mathrm{BH}}}{\pd T}\right)_{\mathrm{sources}}
            =\frac Sz\mathcal S_{\mathrm{BH}}.
        \end{equation}
        The positive sign holds throughout the family \eqref{eq:thermo_family}
        with $S,z>0$.
        Thus the heat capacity is positive for variations of the horizon radius
        at fixed supporting sources. This establishes local thermal stability
        along this one-parameter family. Stability under more general
        thermodynamic or dynamical perturbations requires a separate analysis.

        \section{Massive probe scalar perturbations}
        \label{sec:probe_scalar}

        We consider a free massive probe scalar on the backgrounds constructed
        above, with the metric, gauge field, and supporting dilaton $\phi$ held
        fixed. Its quasinormal spectrum describes scalar relaxation and, in
        separable finite-momentum sectors, dispersion on the anisotropic geometry.
        This gives a first probe of dynamical response relevant to transport and
        of stability against scalar disturbances. Denoting the scalar mass by $m$,
        we use the dimensionless combination $M^2:=m^2/f_0$. Where a holographic description is available, the
        quasinormal frequencies correspond to poles of a scalar retarded
        correlation function~\cite{Son:2002sd,Kovtun:2005ev}.

        \subsection{Klein--Gordon equation on a homogeneous spatial geometry}

        We introduce a real scalar $\Psi$ with probe action
        \begin{equation}
            \label{eq:probe_action}
            I_{\Psi}:=-\frac12\int\dd^{n+2}x\,\sqrt{-g}\,
            \left[g^{\mu\nu}\pd_\mu\Psi\pd_\nu\Psi+m^2\Psi^2\right],
            \qquad (\Box-m^2)\Psi=0.
        \end{equation}
        We work with the metric~\eqref{eq:thermo_family}, although most expressions
        below only require the general metric ansatz~\eqref{eq:metric_ansatz}.
        Its determinant is independent of the blackening factor and obeys
        $\sqrt{-g}=r^{\Delta-1}\sqrt{h_b}$, in the invariant spatial frame. Thus,
        \begin{equation}
            \label{eq:probe_KG_general}
            \left[
                r^{-\Delta+1}\pd_r\bigl(r^{\Delta+1}f\pd_r\bigr)
                -\frac{\pd_t^2}{r^{2z}f}
                +\nabla^2_{h_r}-m^2
            \right]\Psi=0,
            \qquad
            h_r:=\sum_i b_i^2r^{2\alpha_i}(\theta^i)^2.
        \end{equation}
        This expression is valid for any of the neutral or EMD backgrounds considered
        in this paper; the supporting matter fields enter only through the metric.

        The spatial part requires some care. The invariant vector fields $X_i$ introduced above satisfy
        $[X_i,X_j]=c_{ij}{}^kX_k$. The spatial Laplacian takes the form
        \begin{equation}
            \label{eq:probe_spatial_laplacian}
            \nabla^2_{h_r}
            =\sum_i\frac{r^{-2\alpha_i}}{b_i^2}
            \left(X_i^2-\sum_jc_{ij}{}^jX_i\right).
        \end{equation}
        The terms linear in $X_i$ arise from the invariant spatial volume form and
        may be nonzero for non-unimodular Bianchi groups. On a flat horizon,
        $X_i=\pd_i$ and this is the usual Fourier-space Laplacian. For a
        non-Abelian spatial geometry, however, the operators belonging to different
        scaling exponents do not necessarily commute. Therefore a generic spatial
        perturbation cannot always be reduced to a single radial ordinary
        differential equation. One must instead solve a coupled system of spatial
        harmonics, or retain the corresponding radial--spatial partial differential
        equation.

        A universal sector is the spatially homogeneous mode,
        $\Psi(t,r,x)=\e^{-i\omega t}\psi(r)$. At nonzero spatial momentum,
        separation requires an additional condition. Let $\mathcal A$ be the
        set of distinct values among the spatial exponents $\alpha_i$, and define
        \begin{equation}
            L_a:=\sum_{i:\alpha_i=a}\frac1{b_i^2}
            \left(X_i^2-\sum_jc_{ij}{}^jX_i\right),\qquad
            \nabla^2_{h_r}=\sum_{a\in\mathcal A}r^{-2a}L_a.
            \label{eq:probe_grouped_operators}
        \end{equation}
        For a product $\Psi=\e^{-i\omega t}\psi(r)Y(x)$ to reduce the
        equation to a single radial ODE, the same spatial harmonic must obey
        \begin{equation}
            L_aY=-\kappa_aY\qquad\text{for every }a\in\mathcal A.
            \label{eq:probe_joint_harmonics}
        \end{equation}
        The distinct radial powers are independent functions of $r$;
        diagonalizing the spatial Laplacian at one radius is therefore
        insufficient. Operators carrying the same radial power need only be
        diagonalized as a sum, whereas the groups with different powers must
        share the chosen eigenfunction. The $\kappa_a$ are spatial spectral
        labels, distinct from the background parameter $\kappa$.
        With this condition, the equation of motion becomes
        \begin{align}
            \label{eq:probe_radial_eq}
            r^{-\Delta+1}\pd_r\left(r^{\Delta+1}f\psi'\right)
            +\left[\frac{\omega^2}{r^{2z}f}-m^2
            -\sum_{a\in\mathcal A}\kappa_a r^{-2a}\right]\psi&=0,\\
            \left[f\cD^2+(\Delta f+\cD f)\cD
                +\frac{\omega^2}{r^{2z}f}-m^2
                -\sum_{a\in\mathcal A}\kappa_a r^{-2a}\right]\psi&=0.
                \nonumber
        \end{align}
        Here $\cD=r\pd_r$. The homogeneous harmonic has $\kappa_a=0$ for all $a$
        and always satisfies it.

        On a flat horizon, plane waves diagonalize every $\pd_i^2$,
        and $\kappa_a=\sum_{i:\alpha_i=a}k_i^2$, where $k_i$ includes
        the factor $b_i^{-1}$. Only in this case, or another sector admitting
        the corresponding individual eigenfunctions, may the potential be
        interpreted as a sum of independent directional momentum squares.
        On Nil, the equal-scaling horizontal operators instead combine into
        an oscillator operator; its eigenvalue and the central momentum
        yield the separated potential described in
        Sec.~\ref{sec:probe_numerics}. They are not three independent
        Cartesian momenta.

        Solv illustrates the obstruction explicitly:
        \begin{equation}
        \nabla^2_{h_r}=r^{-2}
        (\e^{-2w}\pd_x^2+\e^{2w}\pd_y^2)+b_3^{-2}\pd_w^2.
        \end{equation}
        After Fourier transforming $x,y$, the first group is multiplication
        by $-k_x^2\e^{-2w}-k_y^2\e^{2w}$, while the second differentiates
        in $w$. For generic nonzero transverse momenta these do not share
        a nonzero smooth eigenfunction, so the product ansatz fails.
        Setting $k_x=k_y=0$ leaves separable waves along $w$.

    \subsection{Asymptotic solutions and the scalar mass bound}

    We first consider solutions with the blackening function \eqref{eq:thermo_family},
    $f(r)=f_0-\mu r^{-\Delta}$, for which $f(r)\to f_0>0$ as $r\to\infty$. For the homogeneous mode, the
    frequency-dependent term is subleading when $z>0$ and the asymptotic
    equation becomes
    \begin{equation}
        \label{eq:probe_UV_indicial}
        \left[f_0\cD(\cD+\Delta)-m^2\right]\psi=0.
    \end{equation}
    Define the radial falloff exponents by
    \begin{equation}
        \nu:=\sqrt{\frac{\Delta^2}{4}+M^2},\qquad
        \delta_\pm:=\frac\Delta2\pm\nu.
    \end{equation}
    Away from coincident roots, write $J(\omega)$ and $V(\omega)$ for the
    coefficients of the minus and plus falloffs, respectively. Then
    \begin{equation}
        \label{eq:probe_UV_falloffs}
        \psi(r)\simeq J(\omega)r^{-\delta_-}+V(\omega)r^{-\delta_+}.
    \end{equation}
    This is the direct generalization of the familiar AdS scalar falloffs,
    with the AdS boundary dimension replaced by the effective radial scaling
    exponent $z+S$. In standard Dirichlet quantization, $J$ is the prescribed
    leading boundary datum and $V$ is the response. In a putative dual field
    theory, $\delta_+$ is the corresponding formal scaling dimension in the
    chosen radial convention. The invariant comparison between solutions is
    provided by ratios such as $\delta_+/z$. The holographic identification
    of $J$ and $V$ requires a choice of asymptotic boundary conditions and
    scalar counterterms; it does not follow from the metric alone.

    The roots remain real provided
    \begin{equation}
        \label{eq:probe_BF}
        M^2\geq M_{\mathrm{BF}}^2:=-\frac{\Delta^2}{4}.
    \end{equation}
    This is a BF-like asymptotic mass bound for the homogeneous sector of
    these scaling geometries, analogous to the stability bound at a local
    maximum of AdS \cite{Mezincescu:1984ev}. For planar AdS$_{n+2}$, $z=\alpha_i=1$, $\Delta=n+1$, and
    $f_0=L^{-2}$, recovering $M_{\mathrm{BF}}^2=-(n+1)^2/4$.
    For the five-dimensional neutral Solv solution, one has
    $(z,S,\Delta)=(1,2,3)$ and therefore
    $\delta_\pm=3/2\pm\sqrt{9/4+M^2}$; for Nil,
    $(z,S,\Delta)=(3/2,4,11/2)$ gives
    $\delta_\pm=11/4\pm\sqrt{121/16+M^2}$.
    The radial falloffs are thus sensitive to the anisotropic scaling,
    even for a probe that carries no spatial momentum. Similar massive-scalar
    asymptotics and quasinormal boundary conditions have been examined in
    Lifshitz black holes~\cite{Giacomini:2012hg}.

    There are two conditions for spatially varying modes. First, momentum
    along an unscaled direction, $\alpha_i=0$, is equivalent to a mass shift:
    the associated term $\kappa_0\,r^{0}$ in \eqref{eq:probe_radial_eq}
    survives as $r\to\infty$ and enters the indicial equation at the same
    order as $m^2$. For such a separable mode, with $-L_0Y=k_0^2Y$ and
    $k_0^2\ge0$, the replacement $m^2\to m^2+k_0^2$ must be made in
    \eqref{eq:probe_UV_falloffs}. The neutral Solv background provides
    an example: the third spatial direction is unscaled, so its momentum
    shifts the asymptotic scalar mass. Second, if any $\alpha_i<0$, a
    nonzero spatial derivative in that direction grows toward large $r$.
    In this case momentum may dominate the asymptotic equation, and the
    solutions need not have the radial falloffs of the form~\eqref{eq:probe_UV_falloffs}.
    This issue is relevant, for example, to the Bianchi VI$_h$ family with
    $0<h<1$, where $\alpha_1=-h<0$. The finite-momentum asymptotics must therefore be determined separately
    in these sectors.

    The radial Klein--Gordon norm of the slow falloff behaves asymptotically
    as $\int^\infty\dd r\,r^{-2z-1+2\nu}$. Hence both falloffs are
    normalizable under this elementary norm criterion when $0\leq\nu<z$.
    Normalizability alone does not guarantee that alternative scalar boundary
    conditions define a stable theory. For ordinary Lifshitz backgrounds,
    additional short-wavelength instabilities impose a stronger restriction
    \cite{Andrade:2012xy}; the analogous analysis for the more general
    anisotropic spatial geometries is outside the scope of this paper.

    \subsection{Quasinormal modes}

    For a nonextremal horizon at $r=r_h$, write
    $f(r)=f'(r_h)(r-r_h)+\cdots$. The radial equation gives two local
    solutions, corresponding to ingoing and outgoing waves:
    \begin{equation}
        \label{eq:probe_horizon}
        \psi(r)\sim(r-r_h)^{\mp i\omega/(4\pi T)},
        \qquad
        T=\frac{r_h^{z+1}f'(r_h)}{4\pi}.
    \end{equation}
    With the time dependence $\e^{-i\omega t}$, the minus sign is ingoing
    at the future horizon. Equivalently, the full field is regular in the
    ingoing Eddington--Finkelstein coordinate
    $v:=t+\int\dd r/(r^{z+1}f)$.
    The quasinormal frequencies are those complex $\omega$ for which the
    ingoing solution also obeys the chosen source-free boundary condition,
    $J(\omega)=0$. Under the standard holographic prescription these
    frequencies are poles of the scalar retarded Green function,
    $G_R\propto V/J$, up to contact terms and normalization
    \cite{Son:2002sd, Kovtun:2005ev}. They describe the relaxation of an
    external scalar operator; since $\Psi$ is not a conserved density,
    there is no reason to expect a hydrodynamic mode with $\omega\to0$
    in this probe sector.

    For the blackening function \eqref{eq:thermo_family}, it is useful to introduce
    $u:=r_h/r$ and
    $\mathfrak w:=\omega/(2\pi T)$. \eqref{eq:probe_radial_eq}
    becomes
    \begin{equation}
        \label{eq:probe_compact_eq}
        u^2(1-u^\Delta)\psi_{uu}
        +u(1-\Delta-u^\Delta)\psi_u
        +\left[\frac{\Delta^2\mathfrak w^2 u^{2z}}{4(1-u^\Delta)}
            -M^2
        -\sum_{a\in\mathcal A}\frac{\kappa_a}{f_0r_h^{2a}}u^{2a}\right]\psi=0.
    \end{equation}
    For the spatially homogeneous mode, label the quasinormal overtones by the
    nonnegative integer $N$. Their frequencies have the scaling form
    \begin{equation}
        \label{eq:probe_QNM_scaling}
        \omega_N=2\pi T\,
        \mathfrak w_N\!\left(z,S,M^2\right).
    \end{equation}
    Here $\mathfrak w_N$ is determined by the radial boundary-value problem.
    Thus $\omega_N/T$ is independent of the horizon radius within a fixed
    background of the form \eqref{eq:thermo_family}. For separable finite-momentum modes, the
    additional arguments are $\kappa_a/(f_0r_h^{2a})$, subject to
    the common-harmonic condition~\eqref{eq:probe_joint_harmonics}.
    In general the complex values of
    $\mathfrak w_N$ require a numerical solution, for example by imposing the
    ingoing factor $(1-u)^{-i\omega/(4\pi T)}$ and solving for a vanishing
    leading boundary coefficient.

    \subsection{An analytic quasinormal spectrum for \texorpdfstring{$z=S$}{z=S}}
    \label{sec:probe_analytic}

    A useful simplification occurs for the homogeneous mode when $z=S$,
    so that $\Delta=2z$, provided the blackening factor has the form
    $f=f_0[1-(r_h/r)^{2z}]$, with no logarithmic term. This includes special members of the
    neutral family and does not require the spacetime to be AdS.
    Define
    \begin{equation}
        \label{eq:probe_special_defs}
        y:=\left(\frac{r_h}{r}\right)^{2z},\qquad
        a:=\frac{\delta_+}{2z}
        =\frac12+\frac1{2z}\sqrt{z^2+M^2}.
    \end{equation}
    The radial equation then becomes
    \begin{equation}
        \label{eq:probe_hypergeom_eq}
        y^2(1-y)\psi_{yy}-y^2\psi_y
        +\left[\frac{\mathfrak w^2y}{4(1-y)}
        -\frac{m^2}{4z^2f_0}\right]\psi=0.
    \end{equation}
    This is a hypergeometric equation after extracting its asymptotic
    powers. A solution ingoing at $y=1$ is
    \begin{equation}
        \label{eq:probe_hypergeom_infalling}
        \psi_{\mathrm{in}}=y^a(1-y)^{-i\mathfrak w/2}
        {}_2F_1\!\left(a-i\mathfrak w/2,a-i\mathfrak w/2;
        1-i\mathfrak w;1-y\right).
    \end{equation}
    For $\nu>0$, the coefficient of the slow falloff $y^{1-a}$
    is proportional to
    \begin{equation}
        \label{eq:probe_hypergeom_source}
        J(\omega)\ \propto\
        \frac{\Gamma(1-i\mathfrak w)\Gamma(2a-1)}
        {\Gamma(a-i\mathfrak w/2)^2}.
    \end{equation}
    Vanishing of the source coefficient gives the exact homogeneous
    quasinormal spectrum, with the usual limiting prescription in the
    cases where the arguments of the gamma functions take integer values,
    \begin{equation}
        \label{eq:probe_exact_QNM}
        \omega_N=-i\,4\pi T
        \left(N+\frac{\delta_+}{\Delta}\right),
        \qquad N=0,1,2,\ldots,\qquad z=S.
    \end{equation}
    In particular, for $M^2=0$ one has $\delta_+=\Delta$, and
    $\omega_N=-i4\pi T(N+1)$. These scalar modes are purely damped.
    Since $\delta_+$ is real for any $M^2$ obeying the bound
    \eqref{eq:probe_BF}, the frequencies \eqref{eq:probe_exact_QNM} are
    purely imaginary: at $z=S$ the homogeneous spectrum lies entirely on the
    negative imaginary axis, so the modes are overdamped, with no oscillation
    frequency. Consistently, the time-reversal partners
    $\mathfrak w\to-\bar{\mathfrak w}$ degenerate to self-pairings on the
    axis. For real asymptotic exponents, the spectrum contains no growing homogeneous
    quasinormal modes under the chosen source-free boundary condition. This is
    evidence for stability within this probe sector, not for the coupled
    gravitational and matter perturbations. Away from this homogeneous $z=S$
    sector, the spectrum is determined by the corresponding radial boundary-value
    problem.

    \subsection{Numerical comparison of probe mass and supporting matter}
    \label{sec:probe_numerics}

    We probe the linear response of the anisotropic scaling geometries of
    Sec.~\ref{sec:neutral} with a simple minimally coupled scalar field,
    computing its spectra in the homogeneous and separable finite-momentum
    sectors and varying the probe mass and background parameters under
    standard source-free boundary conditions.  Gravitational quasinormal
    modes of Lifshitz black branes have been computed in
    \cite{Andrade:2022irg}; the scalar sector isolates the linear response
    of the background before the coupled metric, gauge-field, and dilaton
    perturbations are included.

    \paragraph{Background parameters.}
    For fixed spatial exponents, let $z_{\mathrm N}:=Q/S$, with $S>0$.
    The electric extensions of the neutral Nil and Bianchi VI$_h$ solutions
    can be parameterized by
    \begin{align}
        \sigma^2&=S(z-z_{\mathrm N}),\qquad z\geq z_{\mathrm N},\\
        f_0&=\frac{-2\Lambda}{(z+S)[z+(n-1)z_{\mathrm N}]},\qquad
        \cE=(z-z_{\mathrm N})(z+S)f_0,\\
        \lambda\sigma&=-2S,\qquad
        R_{\hat i\hat i}=(\alpha_i-z_{\mathrm N})(z+S)f_0.
        \label{eq:num_background}
    \end{align}
    At the endpoint $z=z_{\mathrm N}$ we use the neutral solution directly.
    Away from it, the spatial curvature scales adjust according to the last
    equation. The scan changes the EMD coupling $\lambda$ and therefore
    compares backgrounds across a family of theories, rather than varying
    an independent charge at fixed coupling. The three families and their
    parameters are listed in Table~\ref{tab:num_backgrounds}. Magnetic and
    dyonic solutions admit the same probe equation below when their
    blackening factor has the same form, but their allowed parameters must
    separately satisfy the magnetic and spatial constraints.

    \begin{table}[htbp]
    \centering
    \begin{tabular}{llll}
    \toprule
    Geometry & Spatial exponents & $S$ & $z_{\mathrm N}$\\
    \midrule
    Nil & $(1,1,2)$ & $4$ & $3/2$\\
    Solv & $(1,1,0)$ & $2$ & $1$\\
    Bianchi VI$_h$ & $(-h,1,0)$ & $1-h$ & $(1+h^2)/(1-h)$\\
    \bottomrule
    \end{tabular}
    \caption{Background families for the homogeneous numerical comparison.
    In the last row $h<1$; Solv is the case $h=-1$.}
    \label{tab:num_backgrounds}
    \end{table}

    \paragraph{Equation and boundary conditions.}
    We use Fourier time dependence
    $\e^{-i\omega v}$ in ingoing Eddington--Finkelstein coordinates,
    $v=t+\int\dd r/(r^{z+1}f)$, with $u=r_h/r$.
    Frequencies are quoted in thermal units,
    \begin{equation}
        \mathfrak w:=\frac{\omega}{2\pi T},\qquad
        2\pi T=\frac{\Delta f_0r_h^z}{2}.
    \end{equation}
    Let $F(u):=f(r_h/u)/f_0=1-u^\Delta$. For each separable harmonic, define
    the momentum potential
    \begin{equation}
        \mathcal K(u):=\sum_{a\in\mathcal A}
        \frac{\kappa_a}{f_0r_h^{2a}}u^{2a}.
    \end{equation}
    It vanishes for the homogeneous mode. When its boundary limit is finite,
    we use the momentum-dependent exponent
    \begin{equation}
        \delta_+:=\frac\Delta2+
        \sqrt{\frac{\Delta^2}{4}+M^2+\mathcal K(0)}.
    \end{equation}
    This reduces to the homogeneous exponent defined above when $\mathcal K(0)=0$.
    For each harmonic write $\Psi=\e^{-i\omega v}u^{\delta_+}y(u)Y(x)$.
    The single numerical equation, including momentum, is
    \begin{align}
    0={}&u^2F y''+
    \left\{u[(2\delta_++1-\Delta)F+uF']
               +i\Delta\mathfrak w u^{z+1}\right\}y'\nonumber\\
    &+\left\{\delta_+[(\delta_+-\Delta)F+uF']
       -M^2-\mathcal K(u)
       +\frac{i\Delta\mathfrak w}{2}(2\delta_+-S)u^z\right\}y.
    \label{eq:num_ef}
    \end{align}
    Primes denote $u$ derivatives. Source-free boundary conditions select
    the finite branch of $y$ at $u=0$, excluding the branch
    $u^{\Delta-2\delta_+}$. In particular $y(0)$ need not vanish.
    At the horizon select the regular ingoing Frobenius branch, satisfying
    \begin{equation}
    \label{eq:num_horizon}
    \begin{aligned}
    0={}&\Delta(-1+i\mathfrak w)y'(1)
    +\left[-\delta_+\Delta-M^2-\mathcal K(1)
    +\frac{i\Delta\mathfrak w}{2}(2\delta_+-S)\right]y(1).
    \end{aligned}
    \end{equation}
    For the numerical calculation, we discretize
    the time-even equation \eqref{eq:probe_compact_eq} in the factorized
    variable
    \begin{equation}
    \label{eq:num_factorized}
        \psi=u^{\delta_+}\,(1-u^\Delta)^{-i\mathfrak w/2}\,q(u),
        \qquad F(u)=1-u^\Delta,
    \end{equation}
    for which the horizon singularity of the $\mathfrak w^2/(1-u^\Delta)$ term
    is removed and the equation becomes a regular nonlinear eigenvalue problem
    on $u\in[0,1]$, with quadratic dependence on $\mathfrak w$,
    \begin{align}
    \label{eq:num_qeq}
        0={}&u^2Fq''+
        \left\{u\bigl[(2\delta_++1-\Delta)-(2\delta_++1)u^\Delta\bigr]
               +i\Delta\mathfrak w\,u^{\Delta+1}\right\}q'\nonumber\\
        &+\left\{\bigl[\mathcal K(0)-\delta_+^2u^\Delta-\mathcal K(u)\bigr]
           +i\Delta\mathfrak w\,\delta_+u^\Delta
           +\frac{\mathfrak w^2\Delta^2}{4}
            \frac{u^{2z}(1-u^{2S})}{1-u^\Delta}\right\}q .
    \end{align}
    This equation is equivalent to \eqref{eq:probe_compact_eq} and was checked
    against the exact hypergeometric solution of
    Sec.~\ref{sec:probe_analytic} at generic complex $\mathfrak w$.
    It is collocated on a Chebyshev--Lobatto grid of $N+1$ points in $u$,
    where $N$ now denotes the collocation resolution rather than an overtone
    number (with the second-derivative matrix built as $D_1^2$).  The horizon
    condition is the row obtained by evaluating \eqref{eq:num_qeq} at $u=1$,
    \begin{equation}
    \label{eq:num_qhorizon}
        \Delta(-1+i\mathfrak w)\,q'(1)
        +\Bigl[\mathcal K(0)-\delta_+^2-\mathcal K(1)
        +i\Delta\mathfrak w\,\delta_+
        +\tfrac12\mathfrak w^2\Delta S\Bigr]q(1)=0 ,
    \end{equation}
    which selects the same regular ingoing branch as
    \eqref{eq:num_horizon}.  The source-free condition is imposed as
    $q'(0)=0$: the homogeneous regular branch behaves as
    $q=q_0+\mathcal O(u^{\min(2z,\Delta)})$ away from resonant expansions,
    and $\min(2z,\Delta)>1$ for the backgrounds scanned here, so the
    divergent branch $u^{\Delta-2\delta_+}(1-u^\Delta)^{i\mathfrak w/2}$ is excluded.
    The resulting nonlinear eigenvalue problem $(\mathbf A+\mathfrak w\mathbf B+\mathfrak w^2
    \mathbf C)\mathbf q=0$ is linearized in the standard companion form and
    solved with the QZ algorithm.  Because the factorized equation
    \eqref{eq:num_qeq} encodes the ingoing branch at the horizon by
    construction, the outgoing (anti-quasinormal) partner of a mode is
    not an eigenvalue of the same problem.
    For real background parameters, the physical spectral symmetry is
    $\mathfrak w\leftrightarrow-\bar{\mathfrak w}$, which preserves the sign
    of the imaginary part.

    Spurious collocation eigenvalues are filtered out by comparing the
    spectra at several resolutions $N$. We retain eigenvalues whose
    relative variation across $N=14,16,18$ is less than $6\times10^{-2}$,
    and report their values at $N=16$. Only damped modes with
    $\operatorname{Im}\mathfrak w<0$ and $|\mathfrak w|\leq12$ are
    included. We call modes passing this resolution-stability test
    \emph{verified}. The filtering tolerance is a selection criterion
    rather than an error estimate; accuracy is checked separately against
    the analytic spectrum and the shooting calculation in
    Appendix~\ref{app:numerics}.

    \paragraph{Parameter scans.}
    We scan $M^2$ and $z$ directly, displaying frequencies as
    $\mathfrak w=\omega/(2\pi T)$. Holding $M^2$ fixed keeps the
    probe mass fixed in radial curvature units. At fixed $\Lambda$, a
    fixed dimensionful mass instead requires using $M^2(z)$.
    The initial comparison uses $M^2=0,1$ and
    $z/z_{\mathrm N}=1,3/2$, followed by continuous scans
    $0\leq M^2\leq4$ and $1\leq(z/z_{\mathrm N})\leq4$; the $M^2=4$
    values appear in the benchmark table of
    Appendix~\ref{app:numerics}.
    The Nil and Solv families provide the principal examples; a fixed
    $h=-1/2$ provides a distinct Bianchi VI$_h$ comparison.
    All these probe masses lie above the homogeneous BF-like bound.

    For each family we compare the massless neutral endpoint with a
    massive neutral probe, a massless probe on the electric extension,
    and a massive probe on that extension. We display both
    $\operatorname{Re}\mathfrak{w}$ and $-\operatorname{Im}\mathfrak{w}$.
    The latter measures damping for a stable mode, and
    $\tau T=[2\pi(-\operatorname{Im}\mathfrak{w})]^{-1}$.
    We track the least damped branches continuously as the parameters vary
    to resolve crossings and changes in their damping order.

    \begin{figure}[htbp]
    \centering
    \includegraphics[width=\linewidth]{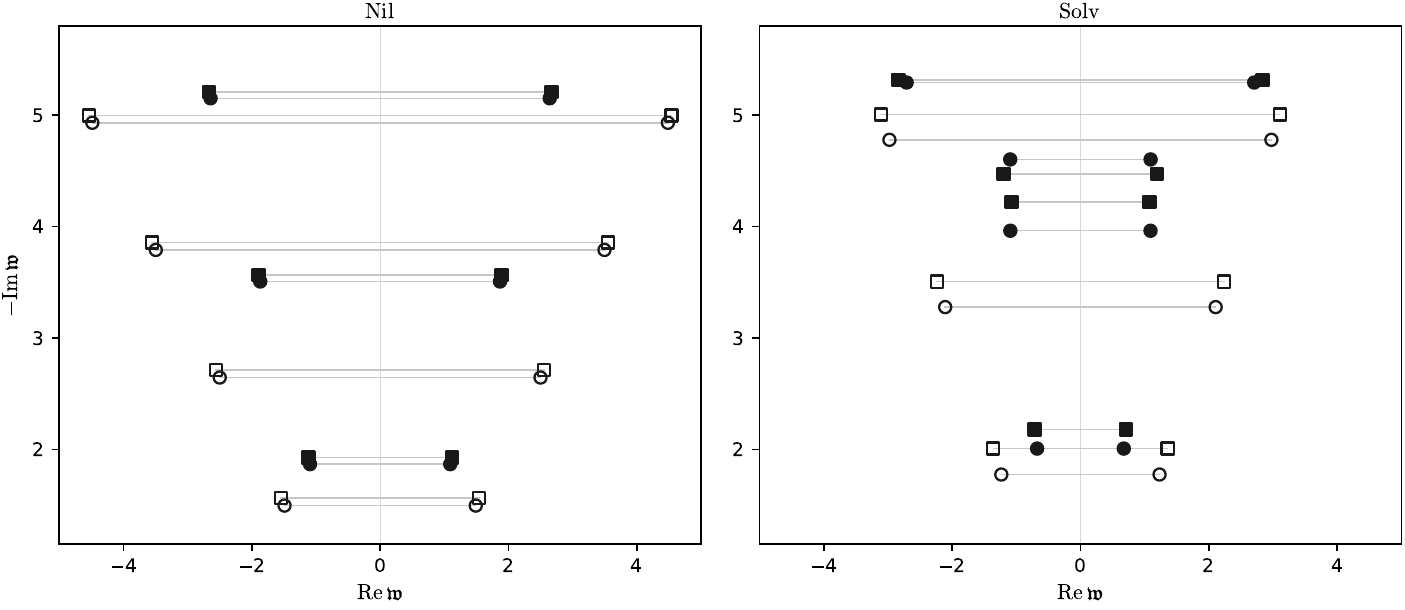}
    \caption{Homogeneous scalar spectra in the complex $\mathfrak w$ plane,
    one panel per family (left: Nil; right: Solv), showing all four
    comparison points $(M^2,z/z_{\mathrm N})=(0,1)$,
    $(1,1)$, $(0,3/2)$ and $(1,3/2)$.  Circles: $M^2=0$;
    squares: $M^2=1$; open markers: the neutral endpoint
    ($z=z_{\mathrm N}$), filled markers: the electrically supported
    extension.  Thin horizontal lines join the two members of each
    $\mathfrak w\leftrightarrow-\bar{\mathfrak w}$ pair.  Modes are
    filtered using $N=22,24,26$ with the same relative tolerance
    ($6\times10^{-2}$) and reported at $N=24$; the panel is restricted to
    the window $|\operatorname{Re}\mathfrak w|\leq5$,
    $-\operatorname{Im}\mathfrak w\leq5.8$.  All numbers quoted in the
    text use $N=14,16,18$.}
    \label{fig:num_spectra}
    \end{figure}

    \begin{figure}[htbp]
    \centering
    \includegraphics[width=\linewidth]{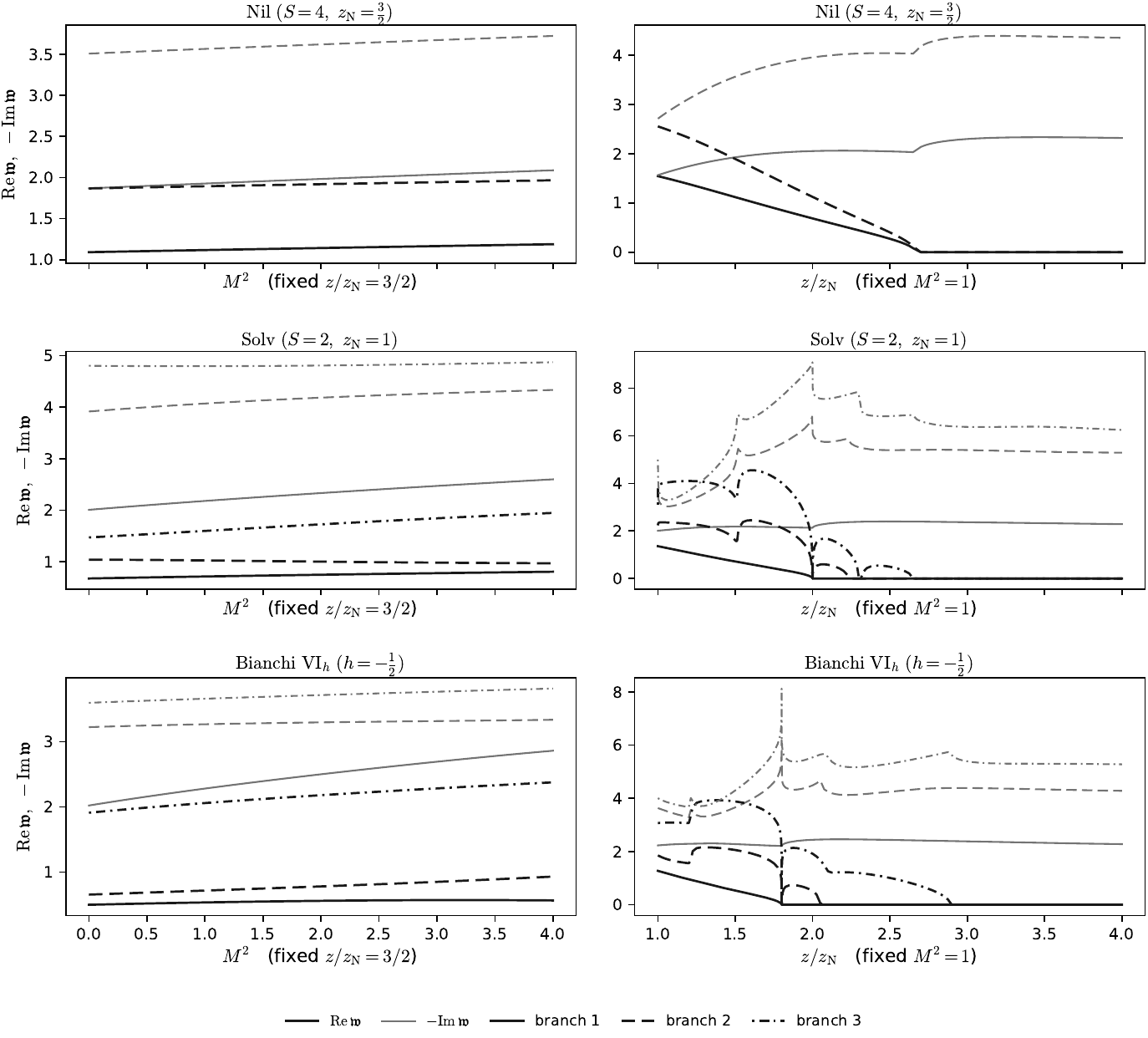}
    \caption{Mass and dynamical-exponent scans of the least damped
    homogeneous scalar branches, for the neutral Nil, Solv and Bianchi
    VI$_h$ ($h=-1/2$) families and their electric extensions.  Left column:
    scan in $M^2$ at fixed $z/z_{\mathrm N}=3/2$; right
    column: scan in $z/z_{\mathrm N}$ at fixed
    $M^2=1$.  In each panel black curves show $\operatorname{Re}\mathfrak w$
    and gray curves $-\operatorname{Im}\mathfrak w$; solid, dashed and
    dash-dotted lines are the fundamental and the first two overtones,
    tracked continuously through the scans.  For each level the member with
    $\operatorname{Re}\mathfrak w\geq0$ is plotted; the partner follows by
    $\mathfrak w\to-\bar{\mathfrak w}$.  Where the scan crosses $z=S$
    (Solv at $z/z_{\mathrm N}=2$, Bianchi VI$_h$ at $z/z_{\mathrm N}=9/5$,
    Nil at $z/z_{\mathrm N}=8/3$), the real parts of a
    pair meet and the modes continue as purely damped levels with split
    damping, in agreement with the analytic spectrum
    \eqref{eq:probe_exact_QNM}.}
    \label{fig:num_scans}
    \end{figure}

    \begin{figure}[htbp]
    \centering
    \includegraphics[width=\linewidth]{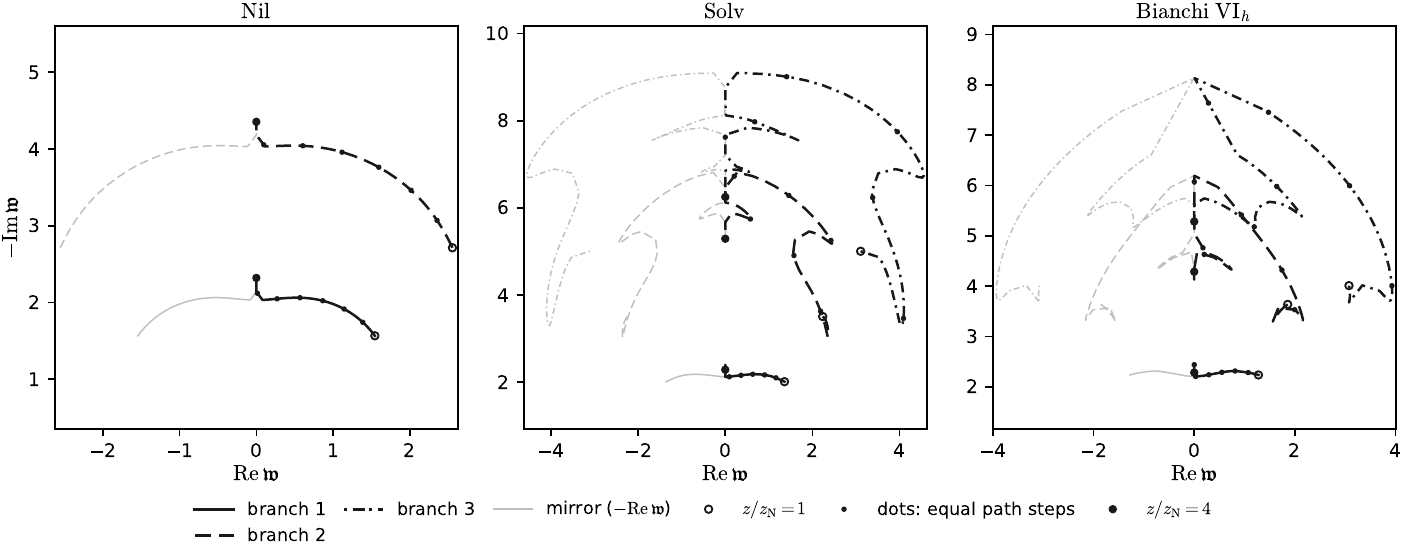}
    \caption{Complex-plane paths of the least damped homogeneous scalar
    branches under the $z/z_{\mathrm N}$ scan at fixed
    $M^2=1$, for the neutral Nil, Solv and Bianchi VI$_h$ ($h=-1/2$)
    families; the same data as the right column of
    Fig.~\ref{fig:num_scans}.  Each branch traces a path as
    $z/z_{\mathrm N}$ increases from $1$ (open circle) to $4$ (filled
    dot); solid, dashed and dash-dotted lines are the fundamental and the
    first two overtones.  The tracked path follows the member with
    $\operatorname{Re}\mathfrak w\geq0$; small dots along each path mark
    equal path-length intervals and do not indicate direction, the open
    circle marks the start of the
    scan ($z/z_{\mathrm N}=1$) and the filled dot its end
    ($z/z_{\mathrm N}=4$), and the mirror path of the
    $\mathfrak w\to-\bar{\mathfrak w}$ partner is drawn as a thin light
    line.  Along each path the real part decreases toward the pinch
    (Solv at $z/z_{\mathrm N}=2$, Bianchi VI$_h$ at $z/z_{\mathrm N}=9/5$,
    Nil at $z/z_{\mathrm N}=8/3$) while the damping rate grows, in
    agreement with the analytic spectrum \eqref{eq:probe_exact_QNM}.}
    \label{fig:num_scans_paths}
    \end{figure}

    \begin{figure}[htbp]
    \centering
    \includegraphics[width=0.60\linewidth]{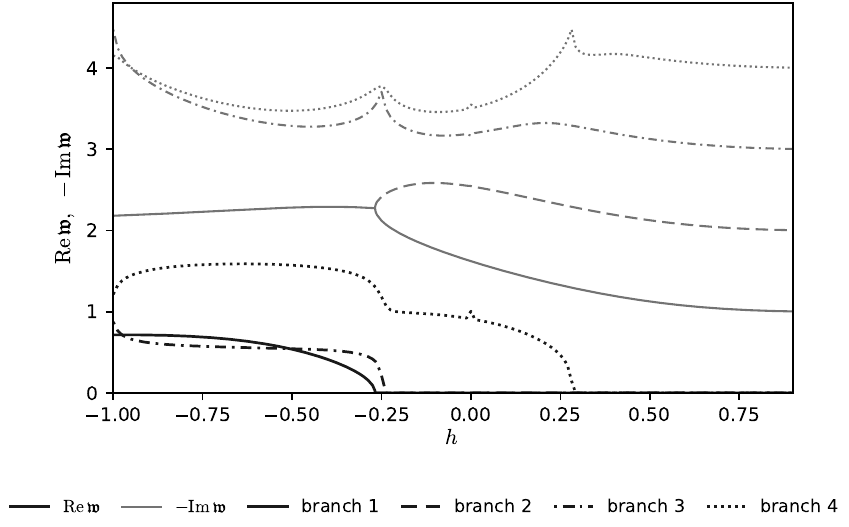}
    \caption{Anisotropy scan of the least damped homogeneous scalar
    branches of the Bianchi VI$_h$ family, at fixed
    $z/z_{\mathrm N}=3/2$ and $M^2=1$ with zero momentum, as $h$ varies
    from $-1$ to $0.9$; the endpoint $h=-1$ is the Solv point of the
    family, and $h=1$, where $z_{\mathrm N}\to\infty$ and the family
    meets a separate Bianchi V branch, is excluded.  Black curves show
    $\operatorname{Re}\mathfrak w$ and gray curves
    $-\operatorname{Im}\mathfrak w$; solid, dashed, dash-dotted and
    dotted lines are the fundamental and the first three overtones,
    tracked continuously through the scan from $h=0$ in both directions.
    For the oscillatory branches the member with
    $\operatorname{Re}\mathfrak w\geq0$ is plotted; the partner follows
    by $\mathfrak w\to-\bar{\mathfrak w}$.  The solid and dashed branches
    meet at the pinch $h\simeq-0.27$, where the scan crosses the $z=S$
    point of the family, and the dashed branch terminates there; the
    kink in the dotted branch at $h\simeq0.28$ is the pinch where its
    oscillatory pair merges on the imaginary axis.}
    \label{fig:num_hscan}
    \end{figure}

    \paragraph{Finite momentum and spatial harmonics.}
    For $-\nabla^2_{h_r}Y=V_k(r)Y$, the momentum potential defined above is
    $\mathcal K(u)=V_k(r_h/u)/f_0$.
    This definition groups all spatial labels into one radial function;
    its finite boundary value is already included in $\delta_+$ above.
    The spatial boundary conditions and harmonic labels must be fixed
    before this reduction; independent Cartesian momenta are generally
    unavailable on a non-Abelian spatial geometry, and a stationary
    brane may carry momentum along the symmetry-breaking directions
    \cite{Iizuka:2014iva}. For compact quotients of these spaces, the
    momenta must belong to the discrete spectrum of the Laplacian on the
    quotient.

    For Nil, the coframe in Sec.~\ref{sec:examples_neutral} gives
    \begin{equation}
    \nabla^2_{h_r}=r^{-2}[\pd_x^2+(\pd_y+x\pd_w)^2]
                   +b_3^{-2}r^{-4}\pd_w^2.
    \end{equation}
    There are two convenient separable sectors. For $k_w=0$, a plane wave
    in $(x,y)$ has
    \begin{equation}
    \widehat k_\perp^2:=\frac{k_x^2+k_y^2}{f_0r_h^2},\qquad
    \mathcal K(u)=\widehat k_\perp^2u^2.
    \label{eq:num_nil_plane}
    \end{equation}
    For $k_w\neq0$, take
    $Y=\e^{i(k_y y+k_w w)}X_\ell(x)$, where
    \begin{equation}
    [-\pd_x^2+(k_y+k_w x)^2]X_\ell
       =(2\ell+1)|k_w|X_\ell,\qquad \ell=0,1,\ldots.
    \end{equation}
    These oscillator harmonics give
    \begin{equation}
    \mathcal K(u)=\frac{(2\ell+1)|k_w|}{f_0r_h^2}u^2
                  +\frac{k_w^2}{f_0b_3^2r_h^4}u^4.
    \label{eq:num_nil_central}
    \end{equation}
    Both momentum terms vanish at the boundary, so the homogeneous
    exponent $\delta_+$ remains valid. The oscillator level is a spatial
    label and is distinct from the radial QNM overtone. At fixed level,
    the limit $k_w\to0$ does not reproduce an arbitrary nonzero transverse
    plane-wave momentum. For the electric Nil family,
    $b_3^2=\Delta f_0$; defining $\rho:=|k_w|/(f_0r_h^2)$ yields
    $\mathcal K=(2\ell+1)\rho u^2+\rho^2u^4/\Delta$.

    For Bianchi VI$_h$, including Solv at $h=-1$, the spatial operator is
    \begin{equation}
    \nabla^2_{h_r}=r^{2h}\e^{-2w}\pd_x^2
        +r^{-2}\e^{-2hw}\pd_y^2
        +b_3^{-2}[\pd_w^2+(1+h)\pd_w].
    \label{eq:num_vi_laplacian}
    \end{equation}
    A separable nonzero-momentum sector is obtained by setting
    $k_x=k_y=0$ and taking
    $Y=\e^{-(1+h)w/2}\e^{ipw}$, $p\in\mathbb R$.
    The prefactor accounts for the measure $\e^{(1+h)w}\dd w$.
    In this sector
    \begin{equation}
    \mathcal K_0:=\frac{p^2+(1+h)^2/4}{f_0b_3^2},\qquad
    \mathcal K(u)=\mathcal K_0.
    \label{eq:num_vi_unscaled}
    \end{equation}
    The radial spectrum therefore depends on $M^2+\mathcal K_0$ exactly
    as the homogeneous spectrum depends on $M^2$. This is a useful numerical
    check. For Solv the measure prefactor disappears and $p=0$ is the
    homogeneous mode. For $h\neq-1$, the constant homogeneous harmonic is
    a separate sector and is not the $p=0$ member of this real-$p$ continuum.

    We restrict the numerical analysis to these separable sectors:
    homogeneous modes, Nil transverse plane waves, Nil central oscillator
    harmonics, and VI$_h$ modes along the unscaled direction. Generic
    transverse momentum on VI$_h$ is outside the scope of the present
    numerical study. Every calculation below therefore reduces to a
    single radial ordinary differential equation.

\begin{figure}[htbp]
    \centering
    \includegraphics[width=\linewidth]{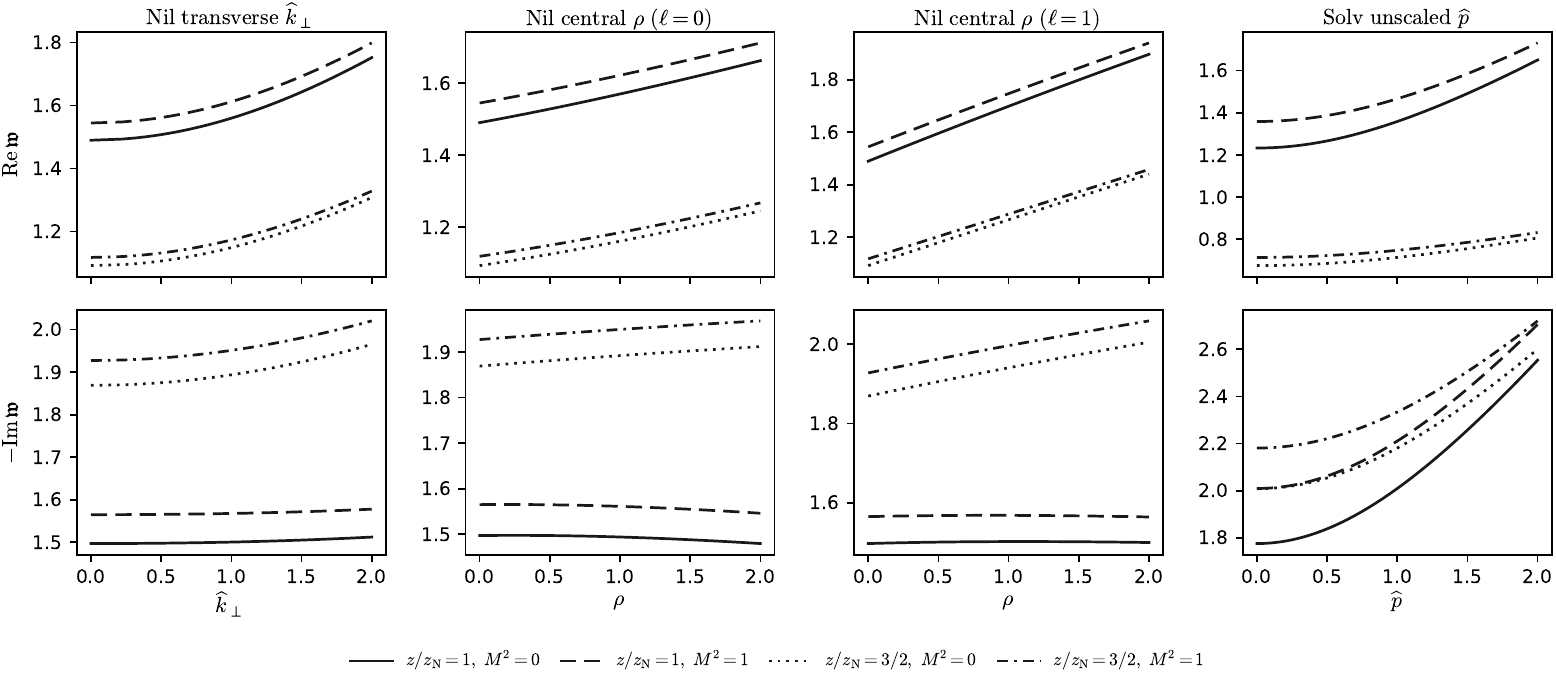}
    \caption{Finite-momentum dispersion of the least damped level (the
    member with $\operatorname{Re}\mathfrak w\geq0$; the partner follows
    by $\mathfrak w\to-\bar{\mathfrak w}$).  Top row:
    $\operatorname{Re}\mathfrak w$; bottom row:
    $-\operatorname{Im}\mathfrak w$.  Columns: Nil transverse plane waves
    \eqref{eq:num_nil_plane}, Nil central oscillator harmonics
    \eqref{eq:num_nil_central} at $\ell=0$ and $\ell=1$, and Solv modes
    along the unscaled direction \eqref{eq:num_vi_unscaled}.  Solid,
    dashed, dotted and dash-dotted lines:
    $(z/z_{\mathrm N},M^2)=(1,0)$, $(1,1)$,
    $(\frac32,0)$, $(\frac32,1)$.  The Solv curves obey the
    effective-mass identification
    $\mathcal M^2\mapsto\mathcal M^2+\widehat p^{\,2}$ exactly
    (verified to machine precision; Appendix~\ref{app:numerics}).}
    \label{fig:num_momentum}
    \end{figure}

    \paragraph{Momentum comparisons.}
    We compute dispersion curves at $M^2=0,1$ and
    $z/z_{\mathrm N}=1,3/2$ for Nil transverse momenta
    $0\leq\widehat k_\perp\leq2$ and Solv unscaled momenta
    $\widehat p:=p/(\sqrt{f_0}b_3)$ in the range $0\leq\widehat p\leq2$.
    For Nil central momentum, we scan $0\leq\rho\leq2$ separately at
    $\ell=0,1$. These dimensionless momenta are held fixed as the background
    parameters vary. Figure~\ref{fig:num_momentum} shows the oscillation
    frequencies and damping rates of the continuously tracked branches,
    with separate panels for the different spatial sectors.

    \paragraph{Validation.}
    The collocation spectra are validated by three independent checks,
    collected in Appendix~\ref{app:numerics}: comparison with the exact
    analytic spectrum at the $z=S$ points of the three families, stability
    of the verified modes under changes of the collocation resolution, and
    an independent shooting calculation. The resolution filter that defines
    the verified modes is a selection criterion rather than an error
    estimate.
    \paragraph{Interpretation.}
    At fixed $(z,S,M^2)$ the homogeneous equation contains no additional
    dependence on the Lie algebra or supporting flux. Any agreement of
    dimensionless spectra at these matched parameters is therefore built
    into the probe equation. The numerical comparison tests how the
    allowed background parameters and probe mass change relaxation; it
    does not independently resolve electric or magnetic charge at fixed
    metric data. Finite spatial momentum and coupled perturbations are
    needed to probe further geometric and matter dependence.

    The computed spectra show the following behavior
    (Figs.~\ref{fig:num_spectra}--\ref{fig:num_momentum}); the spectra,
    benchmark, and momentum results refer to the verified modes, while the
    scans in Figs.~\ref{fig:num_scans} and~\ref{fig:num_scans_paths} track
    the raw $N=16$ spectra to keep the branches continuous through the
    pinch, and the anisotropy scan of Fig.~\ref{fig:num_hscan} tracks the
    raw $N=32$ spectra, since the overtones at $h$ close to $1$ require
    the higher resolution.

    \emph{Mass dependence.}  At fixed $z/z_{\mathrm N}=3/2$,
    raising $M^2$ from $0$ to $4$ increases the damping of the
    least damped branch for all three families while the oscillation
    frequency changes mildly: for the fundamental,
    $(\operatorname{Re}\mathfrak w,-\operatorname{Im}\mathfrak w)$ moves
    from $(1.09,1.87)$ to $(1.19,2.09)$ for Nil, from $(0.68,2.01)$
    to $(0.81,2.60)$ for Solv, and from $(0.50,2.02)$ to
    $(0.56,2.86)$ for Bianchi VI$_h$.  The overtones behave in the same
    direction, so within this family a heavier probe relaxes faster; we do
    not extend this to a stability statement, since the BF-like bound
    \eqref{eq:probe_BF} is only an asymptotic criterion.

    \emph{Dynamical-exponent dependence.}  At fixed $M^2=1$,
    raising $z/z_{\mathrm N}$ from $1$ toward the point $z=S$ of
    each family increases the damping substantially and, for Nil and Solv,
    reduces $|\operatorname{Re}\mathfrak w|$ of the fundamental: from
    $(1.54,1.56)$ at $z/z_{\mathrm N}=1$ to $(0,2.32)$ at
    $z/z_{\mathrm N}=4$ for Nil, and from $(1.36,2.01)$ to
    $(0,2.118)$ for Solv.  The Solv endpoint $z/z_{\mathrm N}=2$
    is the point $z=S=2$, and the fundamental becomes purely damped
    exactly at the analytic value
    $\mathfrak w_0=-2i\,\delta_+/\Delta=-2.11803\,i$; the same pinch
    occurs for Bianchi VI$_h$ at $z/z_{\mathrm N}=9/5$ and for Nil at
    $z/z_{\mathrm N}=8/3$, where
    the two members of the least damped pair meet at
    $\operatorname{Re}\mathfrak w=0$ at the frequencies predicted by
    \eqref{eq:probe_exact_QNM} (for Bianchi VI$_h$:
    $\mathfrak w=-2.20184\,i,\,-4.20184\,i,\ldots$).  This transition of a
    quasinormal pair from oscillatory to purely damped as a state
    parameter is varied, with the pair merging on the imaginary axis and
    splitting into two damped levels, is a familiar feature of
    quasinormal spectra, documented for Schwarzschild--de Sitter
    black holes \cite{Jansen:2017oag} and for holographic
    Einstein--Maxwell--dilaton plasmas, where purely damped modes take
    over as the dominant equilibration channels
    \cite{deOliveira:2024bgh}.  Beyond the pinch the
    two modes remain purely damped with split damping that moves in
    opposite directions: for Bianchi VI$_h$ at $z/z_{\mathrm N}=1.9$
    one finds $-\operatorname{Im}\mathfrak w=1.94$ and $2.40$, for
    Solv at $z/z_{\mathrm N}=5/2$ the values $1.62$ and
    $2.40$ bracket the pre-pinch damping $2.01$, and for Nil at
    $z/z_{\mathrm N}=4$ the values $1.43$ and $2.32$ bracket the
    pinch value $2.03$.  The tracked oscillation
    frequency approaches zero at $z=S$, and the branches were followed
    continuously through this crossing; across $z=S$ each branch continues
    from the corresponding analytic frequency.  The same scans are shown
    as paths in the complex plane in Fig.~\ref{fig:num_scans_paths}.

    \emph{Anisotropy dependence.}  So far the spatial geometry was held
    fixed within each family.  Fig.~\ref{fig:num_hscan} scans the
    anisotropy parameter of the Bianchi VI$_h$ family over
    $-1\leq h\leq0.9$ at fixed $z/z_{\mathrm N}=3/2$ and $M^2=1$, zero
    momentum; the family parameters $S=1-h$ and
    $z_{\mathrm N}=(1+h^2)/(1-h)$ then vary along the scan.  The endpoint
    $h=-1$ is the Solv point of the family, and the fundamental found
    there, $\mathfrak w=\pm0.71-2.18\,i$, is the Solv value of
    Fig.~\ref{fig:num_spectra}; $h=1$, where $z_{\mathrm N}\to\infty$, is
    excluded.  The $z=S$ point of the family moves with $h$ as
    $z/z_{\mathrm N}=(1-h)^2/(1+h^2)$ and crosses the scan value $3/2$
    at $h=\sqrt{3}-2\simeq-0.268$.  For $h$ above the crossing the scan
    point lies beyond the family's pinch and the least damped levels are
    purely damped, with a damping that decreases monotonically with $h$:
    the fundamental moves from $\mathfrak w=-1.62\,i$ at $h=0$ to
    $-1.00\,i$ at $h=0.9$, so at these fixed ratios the brane relaxes
    more slowly as $h$ grows.  At the crossing itself the fundamental
    pair pinches on the imaginary axis exactly at the analytic value
    \eqref{eq:probe_exact_QNM}, $\mathfrak w=-2.27358\,i$ (with
    $z=S=3-\sqrt{3}$), and splits into two purely damped levels whose
    damping moves in opposite directions, bracketing the pinch value
    ($-1.97\,i$ and $-2.53\,i$ at $h=-0.2$); the tracked partner of the
    split pair terminates at the pinch.  For $h$ below the crossing the
    least damped modes are oscillatory, as at the Solv endpoint.  The
    overtones remain oscillatory through the crossing and pinch
    successively at larger $h$ (the dash-dotted branch at $h\simeq-0.24$,
    the dotted branch at $h\simeq0.28$); the remaining oscillatory pairs
    of the window $|\mathfrak w|<12$ pinch at still larger $h$, the last
    ones at $h\simeq0.55$, beyond which the entire window spectrum is
    purely damped.  As $h\to1^-$ the levels arrange themselves into the
    equally spaced ladder $-\operatorname{Im}\mathfrak w=1,2,3,\ldots$: at
    $h=0.9$ the first four levels lie at $1.003$, $2.003$, $3.003$ and
    $4.007$ in these units.

    \emph{Neutral endpoint versus electric extension.}  Comparing the
    neutral endpoint with the electric extension within a family
    (Fig.~\ref{fig:num_spectra}), the extension changes the frequencies
    appreciably already at $z/z_{\mathrm N}=3/2$; since the
    homogeneous equation depends only on
    $(z,S,M^2,\mathcal K)$, this change is controlled by the
    shift of the dynamical exponent $z$ rather than by the supporting flux
    as such.

    \emph{Overtone structure and symmetry.}  The eigenvalues occur in
    pairs $\mathfrak w$ and $-\bar{\mathfrak w}$, visible as the symmetric
    patterns of Fig.~\ref{fig:num_spectra}; at the $z=S$ points the pairs
    collapse onto the imaginary axis and the analytic spectrum
    \eqref{eq:probe_exact_QNM} is equally spaced in damping,
    $\Delta\eta=2$.  Away from $z=S$ the spacing is no longer uniform.

    \emph{Momentum dependence.} For a simple isolated transverse Nil mode,
    the potential depends analytically on $\widehat k_\perp^2$, so the local
    expansion has the form
    $\mathfrak w(\widehat k_\perp)=\mathfrak w(0)
    +c_2\widehat k_\perp^2+O(\widehat k_\perp^4)$,
    with generally complex $c_2$. This expansion need not apply at a mode
    degeneracy. In the central oscillator sector the potential instead
    contains a term linear in $\rho\propto|k_w|$ at fixed $\ell$,
    together with a term quadratic in $\rho$; its small-momentum dependence
    therefore differs from that of transverse plane waves.
    For the unscaled Solv direction, the dispersion is determined by the
    exact effective-mass replacement $M^2\to M^2+\widehat p^2$.

    \section{Conclusions and discussion}
    \label{sec:discussion}

    We have constructed neutral and matter-supported homogeneous anisotropic black branes in arbitrary dimension using the diagonal metric ansatz \eqref{eq:metric_ansatz}, whose spatial coefficients have the form $b_i^2r^{2\alpha_i}$. The spatial Lie algebra enters through scaling compatibility, intrinsic curvature, and the mixed Einstein equations. Together with the radial equations, these conditions determine which homogeneous geometries admit the ansatz. The neutral construction includes the negative-cosmological-constant branch with $\kappa=0$ and a Ricci-flat branch with $\kappa\neq0$, and relates the former to earlier homogeneous black holes and algebraic Ricci solitons. The Ricci-flat branch realizes the hyperscaling-violating vacuum geometries constructed in \cite{Hassaine:2015ifa,Amano:2026ayj}.

    In EMD theory, the electric, magnetic, and dyonic equations retain this separation between radial evolution and geometric constraints. Homogeneous magnetic fields must also satisfy the Maxwell equation and Bianchi identity, so an admissible magnetic orientation cannot be inferred from the Einstein equations alone. We gave electric extensions of five-dimensional Bianchi backgrounds, magnetic Bianchi I and Solv solutions, and a single-$U(1)$ dyonic Nil solution. Higher-dimensional examples include flat dyonic geometries and a family obtained by adjoining a magnetic two-plane to a Ricci nilsoliton. In these solutions, the electromagnetic amplitudes are tied to the scaling background. Independently charged deformations require additional branches beyond those constructed here.

    For $f(r)=f_0[1-(r_h/r)^{z+S}]$, with $f_0,z,S>0$, the horizon quantities give $\mathcal S_{\mathrm{BH}}\propto T^{S/z}$. The Euclidean action at fixed supporting fluxes determines the relative free energy, and the associated mass agrees with the fixed-source Hamiltonian variation. The heat capacity is positive along this one-parameter family. These results do not establish stability under arbitrary source variations or a complete phase diagram. We also do not consider the interior structure of the anisotropic geometries; related static anisotropic black holes have recently been shown to exclude inner horizons and to develop spacelike Kasner-like singularities \cite{Peng:2026baq}. In particular, the electric potential and the exterior electromagnetic energies require a subtraction or renormalization prescription before they can be interpreted as finite boundary quantities.

    The probe scalar gives a first view of wave propagation, dispersion, and relaxation on these backgrounds, as well as a restricted test of stability. Its homogeneous asymptotics give the BF-like bound $M^2\geq-(z+S)^2/4$, which constrains the asymptotic exponents but does not prove global stability. For fixed background parameters, homogeneous quasinormal frequencies are proportional to temperature. At $z=S$, with no logarithmic term in the blackening function, the exact spectrum is
    $\omega_N=-i4\pi T(N+\delta_+/\Delta)$.
    For real asymptotic exponents, the purely damped analytic spectrum \eqref{eq:probe_exact_QNM} supports stability in the homogeneous scalar sector under the chosen boundary conditions and provides an analytic benchmark for the numerical calculation.

    The numerical study extends the homogeneous spectrum to electric Nil, Solv, and Bianchi VI$_h$ backgrounds away from $z=S$. In the Nil, Solv, and Bianchi VI$_h$ scans, the tracked low-lying oscillatory branches meet on the imaginary axis near $z=S$ and continue as purely damped levels. The damping increases with probe mass at the background parameters examined. An anisotropy scan of the Bianchi VI$_h$ family from its Solv point ($h=-1$) toward $h=1$ crosses the $z=S$ point of the family, where the fundamental pair pinches exactly at the analytic frequency; beyond the crossing the least damped levels are purely damped with a damping that decreases as $h$ grows, the oscillatory overtones pinch away successively, and the spectrum approaches an equally spaced purely damped ladder as $h\to1^-$. Finite-momentum calculations also distinguish the spatial geometries: transverse Nil modes have a potential quadratic in transverse momentum, central Nil harmonics involve an oscillator spectrum depending on the absolute central momentum, and momentum along the unscaled Solv direction gives an effective-mass shift. The analytic benchmark, resolution comparisons, and shooting calculation assess the numerical accuracy, while the higher overtones remain more sensitive to resolution. The numerical procedure selects damped branches and therefore cannot rule out growing scalar modes. These results describe scalar dispersion and relaxation in selected sectors, rather than a complete stability analysis or a determination of transport coefficients.

    A natural next step is to study more general, coupled gravitational, dilaton, and $U(1)$ gauge-field perturbations, including inhomogeneous sectors. These would test the dynamical stability of the backgrounds beyond the probe scalar and, where a suitable holographic description is available, reveal the dispersion of conserved-current and stress-tensor fluctuations. The master-function methods of \cite{Kodama:2003kk,Kodama:2007ph} and the gravitational quasinormal-mode literature \cite{Berti:2009kk} provide useful tools where applicable. Such an analysis could also identify hydrodynamic modes and anisotropic transport coefficients.

    Another direction is to relax the purely radial electric ansatz and allow transverse electric-field components along the homogeneous spatial directions. Solving their Maxwell and Bianchi constraints together with the mixed Einstein equations, possibly with a more general stationary metric, would allow us to test whether the enlarged ansatz admits branches carrying nonzero, independently variable electric charge. Such a charge must be identified through the radial electric flux; a transverse electric field does not by itself imply a nonzero conserved charge. A systematic classification of admissible spatial Lie algebras and magnetic two-forms could extend the families found here, while stationary metrics would allow rotation to be included, as in rotating Nil and Solv solutions \cite{Figueroa:2021apr,Faedo:2022hle}.

    For holographic applications, two additional steps are needed: a renormalized variational problem allowing general boundary sources, as developed for Lifshitz asymptotics in \cite{Andrade:2013wsa}, and solutions connecting the homogeneous scaling regions to an appropriate ultraviolet geometry. These would provide the basis for calculating boundary stress tensors, comparing phases in specified ensembles, and studying anisotropic transport.

    \appendix

    \section{Radial integration and logarithmic cases}\label{app:radial}

    For completeness, consider the operator
    \begin{equation}
        P(\cD):=(\cD+2z)(\cD+\Delta),\qquad \cD=r\frac{\dd}{\dd r}.
    \end{equation}
    The neutral and EMD temporal equations have the common form $P(\cD)f=\sum_sJ_s r^s$. Source terms with the same power must first be combined. For a single source the particular solution is
    \begin{equation}
\label{eq:resonance_app}
\begin{aligned}
        f_s(r)=\begin{cases}
            J_s r^s/P(s),&P(s)\neq0,\\[0.3em]
            J_s r^s\log r/P'(s),&P(s)=0,\;P'(s)\neq0,\\[0.3em]
            \tfrac12 J_s r^s(\log r)^2,&P(s)=P'(s)=0.
        \end{cases}
    \end{aligned}
\end{equation}
    The homogeneous solutions are $r^{-2z}$ and $r^{-\Delta}$ when $z\neq S$. For $z=S$ the second homogeneous solution is $r^{-2z}\log r$. In the dyonic equations the source pairs $(s,J_s)$ are
    \begin{equation}
        (0,-4\Lambda/n),\qquad
        (-p,2(n-1)\cE/n),\qquad
        (m_B,2\cB/n).
    \end{equation}
    This covers every case in which a source power coincides with a homogeneous power of the temporal equation, including the case where the electric and magnetic source powers coincide with each other. After integrating, the spatial and dilaton equations must be imposed separately. They can remove an integration constant or rule out a putative logarithmic solution. This is particularly important in the $\Lambda=0$ cases.

    \section{Fixed-source Hamiltonian variation}\label{app:mass}

    We briefly explain the relative mass used in Sec.~\ref{sec:thermo}. For the metric~\eqref{eq:thermo_family}, the radial logarithmic derivative of the volume of a constant-$r$ spatial section is $S/r$. A variation of the blackening mode at fixed spatial scales and fixed EMD supporting sources gives the gravitational Hamiltonian surface variation
    \begin{equation}
        \delta H_{\pd_t}^{\mathrm{grav}}=-\frac{\cV_b S}{2}r^{z+S}\delta f.
    \end{equation}
    Using $\delta f=-r^{-\Delta}\delta\mu$ gives $\delta M=\cV_b S\,\delta\mu/2$, consistent with~\eqref{eq:th_M}. The scalar contribution vanishes because $\delta\phi=0$, and the electric variation vanishes in the fixed-flux ensemble. This derivation specifies the relative variation in the restricted family. It does not determine a universal mass for more general asymptotically homogeneous boundary conditions, nor does it provide the counterterms needed when the matter sources are allowed to vary.

    \section{Numerical convergence and reliability}\label{app:numerics}

    This appendix documents the validation of the collocation spectra of
    Sec.~\ref{sec:probe_numerics}. The accuracy is assessed by comparison
    with the analytic spectrum, by variation of the collocation resolution,
    and by an independent shooting calculation.

\begin{figure}[htbp]
    \centering
    \includegraphics[width=0.62\linewidth]{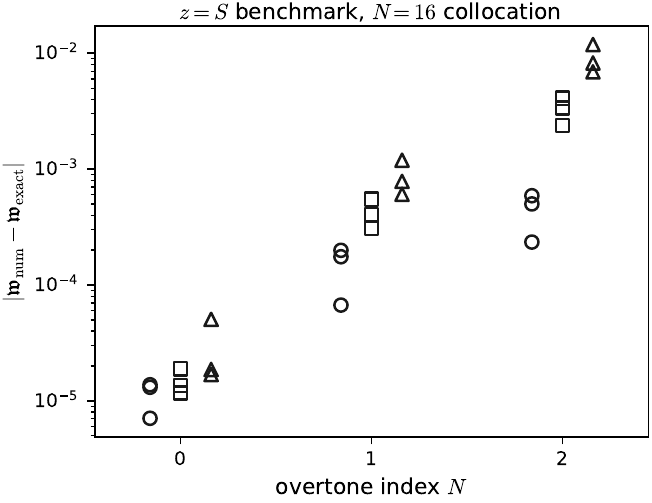}
    \caption{Deviations of the verified collocation modes from the
    analytic spectrum \eqref{eq:probe_exact_QNM} at the three $z=S$ benchmark
    points (open circles: Nil; open squares: Solv; open triangles: Bianchi
    VI$_h$; horizontally offset for readability).  All levels shown are
    verified by the resolution comparison.
    }
    \label{fig:num_validation}
    \end{figure}

    \emph{(i) Analytic benchmark at $z=S$.}  Each electric family admits a
    point $z=S$, where the spectrum
    \eqref{eq:probe_exact_QNM} is exact.  Table~\ref{tab:num_benchmark}
    compares the verified collocation modes with the analytic values at the
    three $z=S$ points (Nil at $z/z_{\mathrm N}=8/3$, Solv at
    $z/z_{\mathrm N}=2$, Bianchi VI$_h$ at
    $z/z_{\mathrm N}=9/5$) for
    $M^2=0,1,4$. For each overtone, let $\mathfrak w_N^{\rm num}$ be the
    numerical dimensionless frequency and define its damping rate by
    $\eta_N^{\rm num}:=-\operatorname{Im}\mathfrak w_N^{\rm num}$. The exact
    damping rates are $\eta_N:=-\operatorname{Im}\mathfrak w_N
    =2(N+\delta_+/\Delta)$. The damping-rate errors are at most
    $1.4\times10^{-5}$ for the fundamental, $7.3\times10^{-4}$
    for the first overtone, and $1.2\times10^{-2}$ for the second
    overtone; the deviations are displayed in
    Fig.~\ref{fig:num_validation}. The table separately reports the maximum absolute error
    in the full complex frequency.
    The
    coincident gamma-function poles do not alter the spectrum in this range:
    for $M^2\geq0$ the numerator pole of the source coefficient
    \eqref{eq:probe_hypergeom_source} is of lower order than the denominator
    poles, so the zeros are unaffected; the massless case
    ($M^2=0$), where the coincident poles occur, is
    reproduced by the numerics at the level quoted above.

    \begin{table}[htbp]
    \centering
    \small
    \begin{tabular}{llcccc}
    \toprule
    Geometry & $M^2$ & $\eta_0^{\rm num}$ &
    $\eta_1^{\rm num}$ & $\eta_2^{\rm num}$ & max dev.\\
    \midrule
    Nil ($z/z_{\mathrm N}=8/3$) & $0$ & $2.00000$ & $3.99999$ & $5.99989$ & $2.3\times10^{-4}$\\
     & $1$ & $2.03077$ & $4.03073$ & $6.03087$ & $5.0\times10^{-4}$\\
     & $4$ & $2.11802$ & $4.11783$ & $6.11744$ & $5.9\times10^{-4}$\\
    Solv ($z/z_{\mathrm N}=2$) & $0$ & $2.00000$ & $3.99985$ & $5.99861$ & $2.4\times10^{-3}$\\
     & $1$ & $2.11803$ & $4.11796$ & $6.11695$ & $4.1\times10^{-3}$\\
     & $4$ & $2.41420$ & $4.41385$ & $6.41100$ & $3.4\times10^{-3}$\\
    VI$_h$ ($h=-\frac12$, $z/z_{\mathrm N}=9/5$) & $0$ & $1.99999$ & $3.99951$ & $5.99446$ & $6.9\times10^{-3}$\\
     & $1$ & $2.20184$ & $4.20112$ & $6.19010$ & $1.2\times10^{-2}$\\
     & $4$ & $2.66667$ & $4.66655$ & $6.66453$ & $8.2\times10^{-3}$\\
    \bottomrule
    \end{tabular}
    \caption{Benchmark at the $z=S$ points: verified collocation frequencies
    $\eta_N^{\rm num}$ compared with the exact damping rates
    $\eta_N=2\bigl(N+\frac{\delta_+}{\Delta}\bigr)$ of
    \eqref{eq:probe_exact_QNM}.  The reported mode is the value at
    $N_{\rm col}=16$. The column ``max dev.'' is the maximum over levels
    $N=0,1,2$ of the absolute complex deviation
    $|\mathfrak w^{\rm num}-\mathfrak w^{\rm exact}|$ at the middle
    resolution, computed from the unrounded benchmark data and includes any
    residual real part of the numerical frequency. Deviations of the damping
    rates are $\leq1.4\times10^{-5}$ ($N=0$), $\leq7.3\times10^{-4}$
    ($N=1$) and $\leq1.2\times10^{-2}$ ($N=2$); all levels shown are
    verified by the filter.}
    \label{tab:num_benchmark}
    \end{table}

    \emph{(ii) Resolution stability.}  Verified modes were required to be
    stable across the three resolutions used in the filter; at a
    representative non-analytic point (Nil electric,
    $z/z_{\mathrm N}=3/2$, $M^2=1$) the verified
    frequencies agree across the resolution sets
    $(12,14,16)$, $(14,16,18)$, $(16,18,20)$ and $(20,22,24)$
    to the quoted digits (identifying the two members of each
    $\mathfrak w\leftrightarrow-\bar{\mathfrak w}$ pair,
    Table~\ref{tab:num_resolution}): the fundamental
    is stable to $3\times10^{-6}$ across the sets, the first overtone
    to $1.4\times10^{-3}$ (Table~\ref{tab:num_resolution}).

    The tracked branches in Fig.~\ref{fig:num_scans} were also compared
    with the filtered spectra at representative points on both sides of
    $z=S$, using $N_{\rm col}=14,16,18$. At coalescence and beyond,
    comparison with $N_{\rm col}=22,24,26$ gives agreement within
    $10^{-4}$ for the fundamental. The pre-coalescence overtones remain
    resolution-limited at the tracking resolution. Adjacent to $z=S$,
    the tracked frequencies approach the analytic values in \eqref{eq:probe_exact_QNM} within
    $2\times10^{-5}$ for the fundamental and $2\times10^{-2}$ for the
    overtones. These comparisons test resolution dependence and branch
    assignment within the collocation method. The branches in
    Fig.~\ref{fig:num_scans} were followed continuously by proximity
    matching on the raw collocation spectra, with adaptive step refinement
    as the overtones rapidly approach the analytic spectrum
    \eqref{eq:probe_exact_QNM}; across $z=S$ each branch continues from
    the corresponding analytic frequency.

    \begin{table}[htbp]
    \centering
    \small
    \begin{tabular}{lcc}
    \toprule
    $N$-triple & fundamental & first overtone\\
    \midrule
    $12,14,16$ & $1.11819 - 1.92737\,i$ & $1.89368 - 3.56570\,i$\\
    $14,16,18$ & $1.11819 - 1.92737\,i$ & $1.89424 - 3.56473\,i$\\
    $16,18,20$ & $1.11819 - 1.92737\,i$ & $1.89430 - 3.56463\,i$\\
    $20,22,24$ & $1.11819 - 1.92737\,i$ & $1.89432 - 3.56446\,i$\\
    \bottomrule
    \end{tabular}
    \caption{Resolution stability of the verified modes at the Nil
    electric point $z/z_{\mathrm N}=3/2$, $M^2=1$: the positive-Re member
    of each $\mathfrak w\leftrightarrow-\bar{\mathfrak w}$ pair at the
    four resolution sets used in the check (the partner follows by
    $\mathfrak w\to-\bar{\mathfrak w}$).  The fundamental varies by
    $3\times10^{-6}$ across the sets, the first overtone by
    $1.4\times10^{-3}$.}
    \label{tab:num_resolution}
    \end{table}

    \emph{(iii) Shooting comparison.} A shooting calculation integrates
    the ingoing horizon Frobenius solution and matches it to the source-free
    boundary branch. Table~\ref{tab:num_shooting} compares the imaginary
    parts of the fundamental frequencies. For the five entries other than
    the VI$_h$ neutral endpoint, the differences calculated from the displayed
    values range from $1.0\times10^{-5}$ to $2.45\times10^{-3}$.
    The shooting root search at the VI$_h$ neutral endpoint did not satisfy
    its convergence tolerance; that entry is reported for reference and is
    excluded from the validation range. Agreement of the imaginary parts
    alone does not determine the error in the full complex frequency.

    \begin{table}[htbp]
    \centering
    \small
    \begin{tabular}{llcc}
    \toprule
    Geometry & $z/z_{\mathrm N}$ & $\operatorname{Im}\mathfrak w_{\rm coll}$ & $\operatorname{Im}\mathfrak w_{\rm shoot}$\\
    \midrule
    Nil & $1$ & $-1.56499$ & $-1.56500$\\
    Nil & $3/2$ & $-1.92737$ & $-1.92757$\\
    Solv & $1$ & $-2.00944$ & $-2.00699$\\
    Solv & $3/2$ & $-2.18087$ & $-2.17947$\\
    VI$_h$, $h=-\frac12$ & $1$ & $-2.23266$ & $-2.23198^{\dagger}$\\
    VI$_h$, $h=-\frac12$ & $3/2$ & $-2.28228$ & $-2.28390$\\
    \bottomrule
    \end{tabular}
    \caption{Imaginary parts of the fundamental frequencies at $M^2=1$
    from collocation and shooting. Negative imaginary parts indicate damping;
    the real parts are not displayed. The VI$_h$ neutral-endpoint shooting
    result (marked $\dagger$) did not converge within the root-finding
    tolerance and is excluded from the accuracy comparison.}
    \label{tab:num_shooting}
    \end{table}

    An additional exact check uses the Solv unscaled-direction sector: since
    its potential is constant, $\mathcal K_0=\widehat p^{\,2}$, the radial
    spectrum at $(\widehat p,M^2)$ must coincide with the
    homogeneous spectrum at $(0,M^2+\widehat p^{\,2})$.  The
    collocation modes satisfy this identification to machine precision for
    all tested combinations.

    \bibliographystyle{unsrturl}
    \bibliography{references}
    \end{document}